\documentclass[aps,prl,twocolumn,superscriptaddress]{revtex4-1} 
\usepackage{graphicx,subfigure}
\usepackage{amsmath,amssymb,bm}
\usepackage{mathtools}
\usepackage{multirow}
\usepackage{textcomp}
\usepackage{float}
\usepackage{color}
\usepackage{ulem}
\usepackage{dsfont}
\usepackage{mathrsfs}
\usepackage{pdfpages}

\renewcommand{\vec}[1]{\ensuremath{\mathbf{#1}}}
\newcommand{\uvec}[1]{\ensuremath{\hat{\mathbf{#1}}}}
\newcommand{\abs}[1]{\ensuremath{\vert{#1}\vert}}
\newcommand{\avg}[1]{\langle #1 \rangle}
\newcommand{\ket}[1]{\ensuremath{\left\vert{#1}\right\rangle}}

\newcommand{\Htot}{H_\mathrm{tot}}
\newcommand{\sF}{\mathcal{F}}
\newcommand{\w}{\ensuremath{c}}
\renewcommand{\th}{\ensuremath{\Theta}}

\newcommand{\avgOmega}{\ensuremath{\Omega}}
\newcommand{\n}{\ensuremath{n}}

\newcommand{\micro}[1]{\ensuremath{\mu\mathrm{#1}}}

\makeatletter
\AtBeginDocument{\let\LS@rot\@undefined}
\makeatother

\begin{document}

\title{Protecting Spin Coherence in a Tunable Heisenberg Model}
\author{Emily J. Davis}
\author{Avikar Periwal}
\author{Eric S. Cooper}
\affiliation{Department of Physics, Stanford University, Stanford, California 94305, USA}
\author{Gregory Bentsen}
\affiliation{Department of Physics, Stanford University, Stanford, California 94305, USA}
\affiliation{Department of Physics, Princeton University, Princeton, New Jersey 08540, USA}
\author{Simon J. Evered}
\affiliation{Department of Physics, Stanford University, Stanford, California 94305, USA}
\author{Katherine Van Kirk}
\affiliation{Department of Physics, Stanford University, Stanford, California 94305, USA}
\author{Monika H. Schleier-Smith}
\affiliation{Department of Physics, Stanford University, Stanford, California 94305, USA}
\affiliation{SLAC National Accelerator Laboratory, Menlo Park, CA 94025}
\begin{abstract}
Using an ensemble of atoms in an optical cavity, we engineer a family of nonlocal Heisenberg Hamiltonians with continuously tunable anisotropy of the spin-spin couplings.  We thus gain access to a rich phase diagram, including a paramagnetic-to-ferromagnetic Ising phase transition that manifests as a diverging magnetic susceptibility at the critical point. The susceptibility displays a symmetry between Ising interactions and XY (spin-exchange) interactions of the opposite sign, which is indicative of the spatially extended atomic system behaving as a single collective spin. Images of the magnetization dynamics show that spin-exchange interactions protect the coherence of the collective spin, even against inhomogeneous fields that completely dephase the non-interacting and Ising systems. Our results underscore prospects for harnessing spin-exchange interactions to enhance the robustness of spin squeezing protocols.
\end{abstract}
\date{\today}
\maketitle
Models of quantum magnetism capture the physics of diverse systems ranging from ferromagnets to resonating valence bond solids \cite{anderson1973resonating} and quantum spin liquids \cite{kitaev2006anyons}.  Implementing such models using cold atoms or molecules \cite{duan2003controlling,barnett2006quantum,trotzky2008time, simon2011quantum,browaeys2020many, bernien2017probing, zeiher2017coherent,guardado2018probing} opens pathways both for elucidating the physics of materials and for accessing new quantum many-body phenomena \cite{micheli2006toolbox, glaetzle2015designing,yao2018quantum,hung2016quantum,strack2011dicke, gopalakrishnan2011frustration,buchhold2013dicke, swingle2016measuring, bentsen2019treelike,marino2019cavity,mivehvar2019cavity,bentsen2019integrable,kim2019low,defenu2018dynamical,colella2018quantum}.  Several prospects, including simulating spin glasses \cite{strack2011dicke, gopalakrishnan2011frustration,buchhold2013dicke} or information scrambling in black holes \cite{swingle2016measuring, bentsen2019treelike,marino2019cavity}, require exotic non-local interactions.  Non-local spin-spin couplings can also aid in combinatorial optimization \cite{johnson2011quantum, lucas2014ising}, investigating new forms of integrability \cite{bentsen2019integrable} or non-equilibrium phase transitions \cite{morrison2008dynamical,chiacchio2019dissipation,muniz2019exploring}, and preparing entangled states \cite{leroux2010implementation,hosten2016quantum,braverman2019near,barontini2015deterministic,davis2016approaching,hu2017vacuum,lewis2018robust,masson2017cavity}.


Nonlocal spin models are naturally realized in cavity-QED experiments, where the cavity mediates interactions among distant atoms \cite{leroux2010implementation,hosten2016quantum,braverman2019near,barontini2015deterministic,vaidya2018tunable,norcia2018cavity,landini2018formation,davis2019photon,muniz2019exploring,spethmann2016cavity,georges2018light}.  For atoms coupled to a single cavity mode, the dynamics are often approximated by viewing the system as a single collective spin \cite{kuzmich2004nonsymmetric,hu2015entangled,dellantonio2017multiparticle}.  In practice, spatial inhomogeneities interfere with this approximation. For studies of many-body physics, inhomogeneities facilitate access to a larger Hilbert space compared with that of a spatially uniform system \cite{bentsen2019integrable,marino2019cavity}. In the context of quantum state engineering, however, inhomogeneities adversely impact metrological protocols that benefit from maximizing spin coherence~\cite{hosten2016quantum,hu2015entangled,dellantonio2017multiparticle, wu2019retrieval}. 

A demonstrated approach to protecting spin coherence is to harness suitably designed interactions \cite{deutsch2010spin,buening2011extended,choi2017observation,zhang2017observation}.  Examples include collisional spin self-rephasing in Bose-Einstein condensates \cite{deutsch2010spin,buening2011extended} and temporal ordering in disordered dipolar materials \cite{choi2017observation}. 
In the cavity-QED context, spin-exchange interactions \cite{norcia2018cavity,davis2019photon} have been proposed as a mechanism for preventing dephasing during spin squeezing protocols \cite{lewis2018robust}, by providing an energy gap between manifolds of different total spin \cite{norcia2018cavity}.  While Ref. \cite{norcia2018cavity} has shown spectroscopic evidence of this energy gap, an observation of enhanced spin coherence---or a comparison with Ising interactions employed for squeezing to date \cite{leroux2010implementation,hosten2016quantum,braverman2019near}---has hitherto been lacking.

\begin{figure}[tb]
\includegraphics[width=\columnwidth]{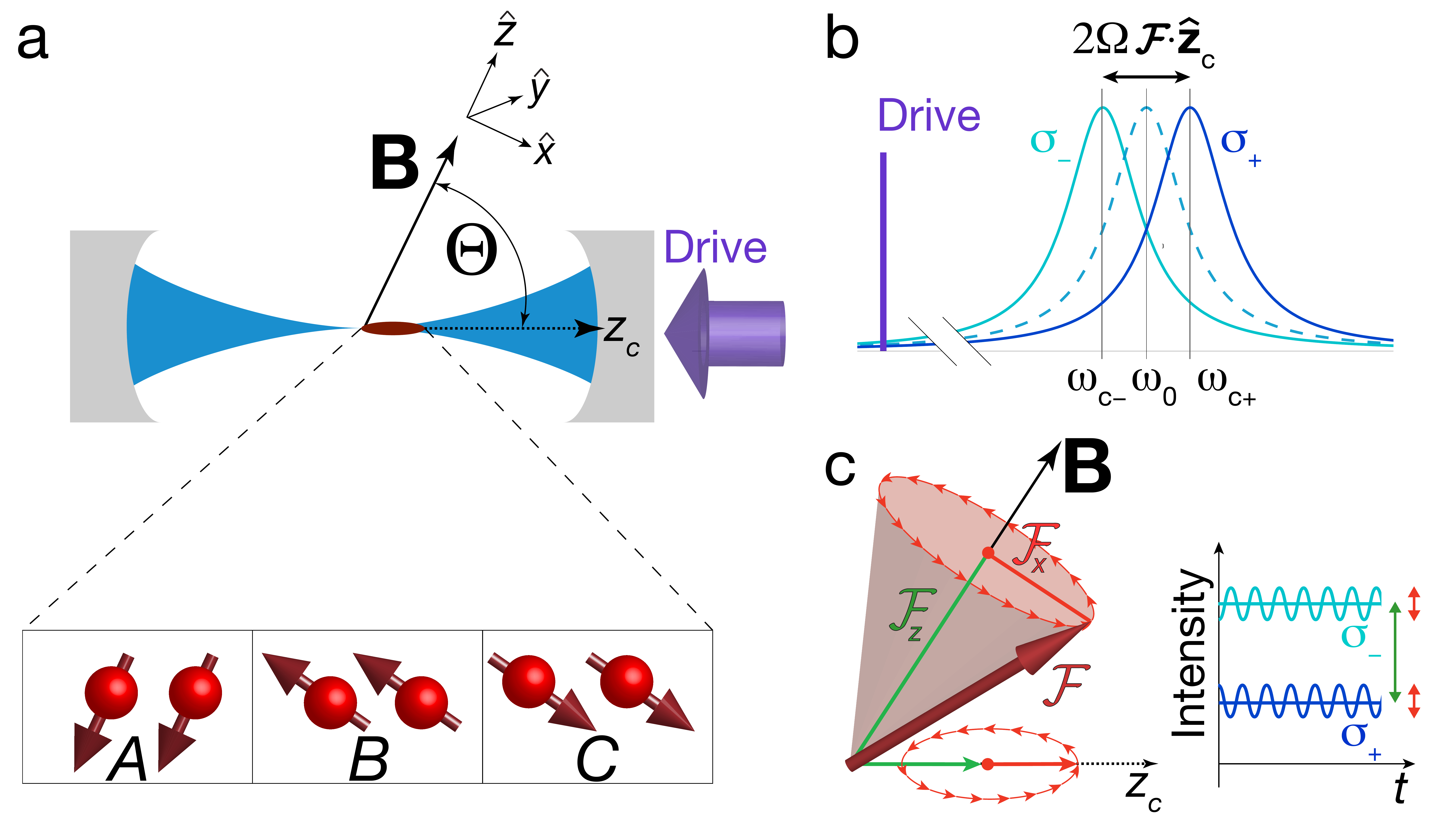}
\caption{Experimental scheme for engineering Heisenberg interactions. (a) Atomic spins in optical cavity precess about magnetic field $\vec{B}$ at angle $\th$ from cavity axis $\uvec{z}_c$. Zoom shows spin texture with three distinct regions $A$, $B$, $C$ as starting point for Hamiltonian tomography. (b) Atom-induced birefringent splitting $\omega_{c+} - \omega_{c-} = 2\Omega \bm{\sF}\cdot \uvec{z}_c$ of $\sigma_{\pm}$ cavity modes.  (c)  Effect of precessing spins on intracavity intensities of $\sigma_{\pm}$ light: projection of static $z$-component (green) shifts DC level to generate Ising interactions, while oscillating projections of transverse components (red) modulate the intensities to generate spin-exchange interactions.
}\label{fig:schematic}
\end{figure}

\begin{figure*}[htb]
\centering
\includegraphics[width=\linewidth]{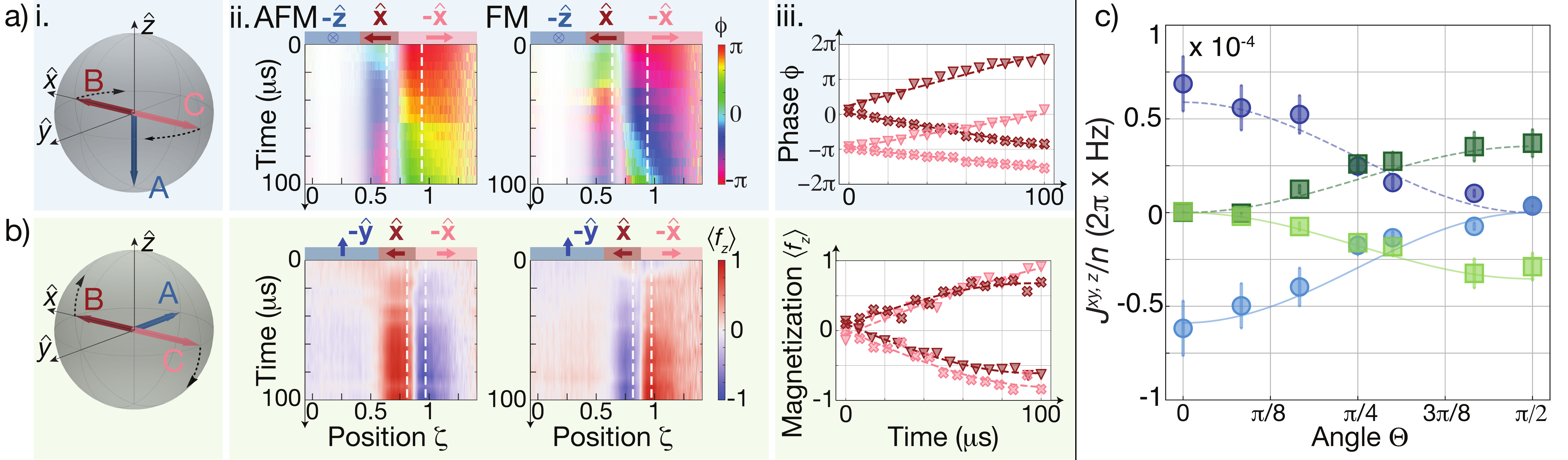}
\caption{
Hamiltonian tomography for determining (a) Ising and (b) XY couplings. (i) Schematic of initial state and its evolution, showing probe spins precessing about mean field along (a) $-\uvec{z}$ or (b) $-\uvec{y}$. (ii) Representative measurements at $\th \approx 53^\circ$ with $\abs{\vec{B}}=3$~G. The initial state is indicated above each plot, showing the direction of the spin vector in regions $A, B, C$. These measurements show (a) phase $\phi$, with opacity indicating transverse spin length, or (b) magnetization $\avg{f_z}$.  We measure with both signs of drive detuning $\delta$ to compare antiferromagnetic (AFM) [(a) $\delta = 2\pi\times 7.5$~MHz, (b) $2\pi\times 5.5$~MHz] and ferromagnetic (FM) [(a) $\delta= -2\pi\times 5.5$~MHz (b) $ -2\pi\times 5.5$~MHz] couplings. (iii) Cuts through the regions initially polarized along $\uvec{x}$ (crimson) and $-\uvec{x}$ (pink) showing (a) $\phi(t)$ with linear fits and (b) $\avg{f_z(t)}$ with sinusoidal fits. Crosses/triangles are for AFM/FM couplings.  (c) $J^z$ (blue circles) and $J^{xy}$ (green squares) vs $\th$. Dark/light markers are for blue/red drive detuning $\delta = \pm 2\pi\times 5.3(4)~$MHz.
}
\label{fig:Figure2}
\end{figure*}

Here, we report on realizing and probing a family of tunable nonlocal Heisenberg models for spins encoded in Zeeman states of atoms in an optical cavity (Fig. 1). The strengths and signs of spin-exchange (XY) and Ising couplings mediated by the cavity are fully controlled by magnetic and optical fields.  We demonstrate this tunability by Hamiltonian tomography and by probing the magnetic susceptibility.  The susceptibility reveals a paramagnetic-to-ferromagnetic phase transition arising in both the \textit{ferromagnetic} Ising and \textit{antiferromagnetic} XY models. Yet comparing effects of Ising and spin-exchange interactions on spin coherence reveals a striking difference, with XY interactions protecting against inhomogeneous fields that otherwise completely dephase the system. 

The interactions that we engineer are described by a Hamiltonian \cite{bentsen2019integrable}
\begin{equation}
\label{eq:H}
H_\mathrm{XXZ}/\hbar = J^{xy}(\th)\left[\sF_x^2 + \sF_y^2\right] + J^z(\th)\sF_z^2.
\end{equation}
Here $\bm{\mathcal{F}} \equiv \sum_i \w_i \vec{f}_i$ is a weighted collective spin vector that accounts for non-uniform couplings to the cavity mode, parameterized by $\w_i$ for atom $i$ with spin $\vec{f}_i$. The weights $c_i$ are normalized such that they average to one.
The relative strength of the spin-exchange coupling $J^{xy}$ and Ising coupling $J^z$ is governed by the angle $\th$ between the cavity axis and an external magnetic field $\vec{B}$, which defines the quantization axis $\uvec{z} = \vec{B}/B$ for the spins [Fig.~\ref{fig:schematic}(a)].  We view the system in a frame rotating about $\vec{B}$ at the Larmor frequency (Zeeman splitting) $\omega_Z = \mu_B B/2$.

The underlying mechanism for the tunable interactions is the Faraday effect \cite{hammerer2010quantum}: the atomic magnetization generates circular birefringence for the intracavity light, which acts back on the atoms via a vector light shift.  Specifically, the  magnetization component along the cavity axis $\uvec{z}_c$ couples to the light, introducing a birefringent splitting $\omega_{c+} - \omega_{c-} = 2\avgOmega \bm{\sF}\cdot \uvec{z}_c$ between the frequencies of the $\sigma_\pm$ cavity modes [Fig. \ref{fig:schematic}(b)], where $\avgOmega$ parameterizes the average birefringence per atom. 
Driving the cavity with linearly polarized light of frequency $\omega_d$ injects $\sigma_+$ and $\sigma_-$ photons into the cavity. For large drive detuning $\delta = \omega_d - \omega_{0}$ from cavity resonance, the birefringence unbalances the $\sigma_\pm$ intracavity intensities [Fig. \ref{fig:schematic}(c)], producing a vector light shift. 

Ising interactions arise when the applied magnetic field $\vec{B}$ is oriented along the cavity axis, i.e., $\uvec{z} = \uvec{z}_c$.  The Faraday effect then yields a vector light shift $\propto \sF_z$, producing the $\sF_z^2$ terms in the Hamiltonian, akin to Refs. \cite{leroux2010implementation,hosten2016quantum,braverman2019near}.  XY interactions arise when $\vec{B}$ has a component orthogonal to the cavity axis \cite{davis2019photon}.  In the lab frame, the transverse spin components ($x,y$) then have oscillating projections along the cavity axis that modulate the polarization of the intracavity light at the Larmor frequency $\omega_Z$ [Fig.~\ref{fig:schematic}c]. This polarization modulation acts as a rotating transverse field that drives spin flips. 
Provided that the drive is detuned from resonance for Raman processes that flip a single spin~\cite{kohler2017cavity}, the lowest-order resonant process is pairwise spin-exchange
\cite{davis2019photon}.

Our experiments employ the $f=1$ hyperfine spins of $^{87}$Rb atoms, which interact via a near-concentric optical cavity (Fig. \ref{fig:schematic}).  A cloud of $N =  1 \times 10^5$ laser-cooled atoms is trapped in a 1560-nm intracavity lattice.  The cloud length is comparable to the Rayleigh range $z_R = 1.4~$mm of the cavity.  Interactions are mediated by a 780-nm TEM$_{00}$ mode of linewidth $\kappa=2\pi\times 200$~kHz, detuned by $\Delta = -2\pi \times 11~$GHz from the $\ket{5S_{1/2}, f=1} \rightarrow \ket{5P_{3/2}}$ transition.  The vacuum Rabi frequency $2g=2\pi\times2.5~$MHz on the cycling transition produces a maximal vector light shift $\Omega_0 = -g^2/6\Delta = 2\pi\times 23~\mathrm{Hz}$
per circularly polarized intracavity photon, for a cold atom at an antinode at cavity center.  This value is reduced to $\avgOmega = 2\pi\times 7(1) ~\mathrm{Hz}$ for an average atom due to the rms transverse cloud size of $13~\micro{m}$ and to displacement from cavity center \cite{SM}.



We benchmark our implementation of the tunable Heisenberg Hamiltonian (Eq.~\ref{eq:H}) by extracting the Ising and XY couplings $J^{z, xy}$ from quench dynamics. We design initial states such that $J^{z}$ or $J^{xy}$ can be transparently extracted from the rate and direction of probe spins in regions $B$ and $C$ precessing about an effective field due to the spins in region $A$ [Fig.~\ref{fig:schematic}(a)]. By scanning a focused Raman beam across the cloud, we prepare initial states of the form $\ket{\psi_\alpha}~=~\ket{\hat{\boldsymbol{\alpha}}}_A\ket{\uvec{x}}_B\ket{-\uvec{x}}_C$, where $\ket{\uvec{u}}_R$ denotes a spin-polarized state along $\uvec{u}$ in region $R$~\cite{SM}. To measure the Ising or XY couplings [Fig. \ref{fig:Figure2}(a-b)], we orient the spins in region $A$ along $\hat{\boldsymbol{\alpha}}=-\uvec{z}$ or $\hat{\boldsymbol{\alpha}}=-\uvec{y}$, respectively. We prepare probe spin vectors $\bm{\sF}^B, \bm{\sF}^C$ that point in opposite directions and are approximately equal in length, such that they ideally produce no net mean field.

After initializing the desired spin texture, we switch on the drive field to induce evolution under $H_\mathrm{XXZ}$. Representative measurements are shown in Fig.~\ref{fig:Figure2}(a-b) for a field angle $\th \approx 53^\circ$, where we expect both the Ising and XY couplings $J^{z,xy}$ to be nonzero. We extract $J^z$ from the phase $\phi \equiv \text{arg}[\avg{f_x} + i \avg{f_y}]$ of spins in regions $B$ and $C$ precessing about $\bm{\mathcal{F}}^A\propto \uvec{z}$ in Fig. \ref{fig:Figure2}a~\cite{SM}. We extract $J^{xy}$ analogously from measurements of the magnetization $\avg{f_z} = \abs{\avg{\vec{f}}}\cos\theta$ of spins in regions $B$ and $C$ rotating about $\bm{\mathcal{F}}^A\propto \uvec{y}$ in Fig.~\ref{fig:Figure2}b.  In each case, we compare red and blue drive detunings $\delta$
and find opposite signs of the spin rotation, indicating opposite signs of interaction \cite{davis2019photon,SM}.  The spatial gradient in rotation rates arises from the dependence of atom-cavity coupling $\w(\zeta)$ on distance $\zeta\equiv z_c/z_R$ from cavity center \cite{SM}.  


The tunability of the interactions via the field angle $\th$ is illustrated in Fig.~\ref{fig:Figure2}c.  For each angle, we obtain the spin-spin couplings $J^{xy,z}$ from fits to the local time evolution $\phi(t)$ or $\theta(t)$ at two positions
with local couplings $\w$, as in Fig.~\ref{fig:Figure2}(a-b.iii).  Specifically, we plot the average spin-spin coupling per intracavity photon $J^z/\n = \dot{\phi}/\n \w\sF_{z}$ (blue circles) and  $J^{xy}/\n~=~\dot{ \theta}\cos\phi /\n\w\sF_{y}$ (green squares), measured with typical intracavity photon number $\n\approx 5000$. 
We fit the data with functional forms $J^z(\th) = J^z(0)\cos^2\th$ and $J^{xy}(\th) = J^{xy}(\pi/2) \sin^2\th$.  The results are consistent with the model of the Faraday interaction, in which the couplings approach $J^z(0)= 2J^{xy}(\pi/2) = \n\avgOmega^2/\delta$ in the large-detuning limit $|\delta| \gg \omega_Z,\kappa$ \cite{SM}. The tomography thus confirms that we have successfully engineered $H_\mathrm{XXZ}$.



\begin{figure}[htb]
\includegraphics[width=\columnwidth]{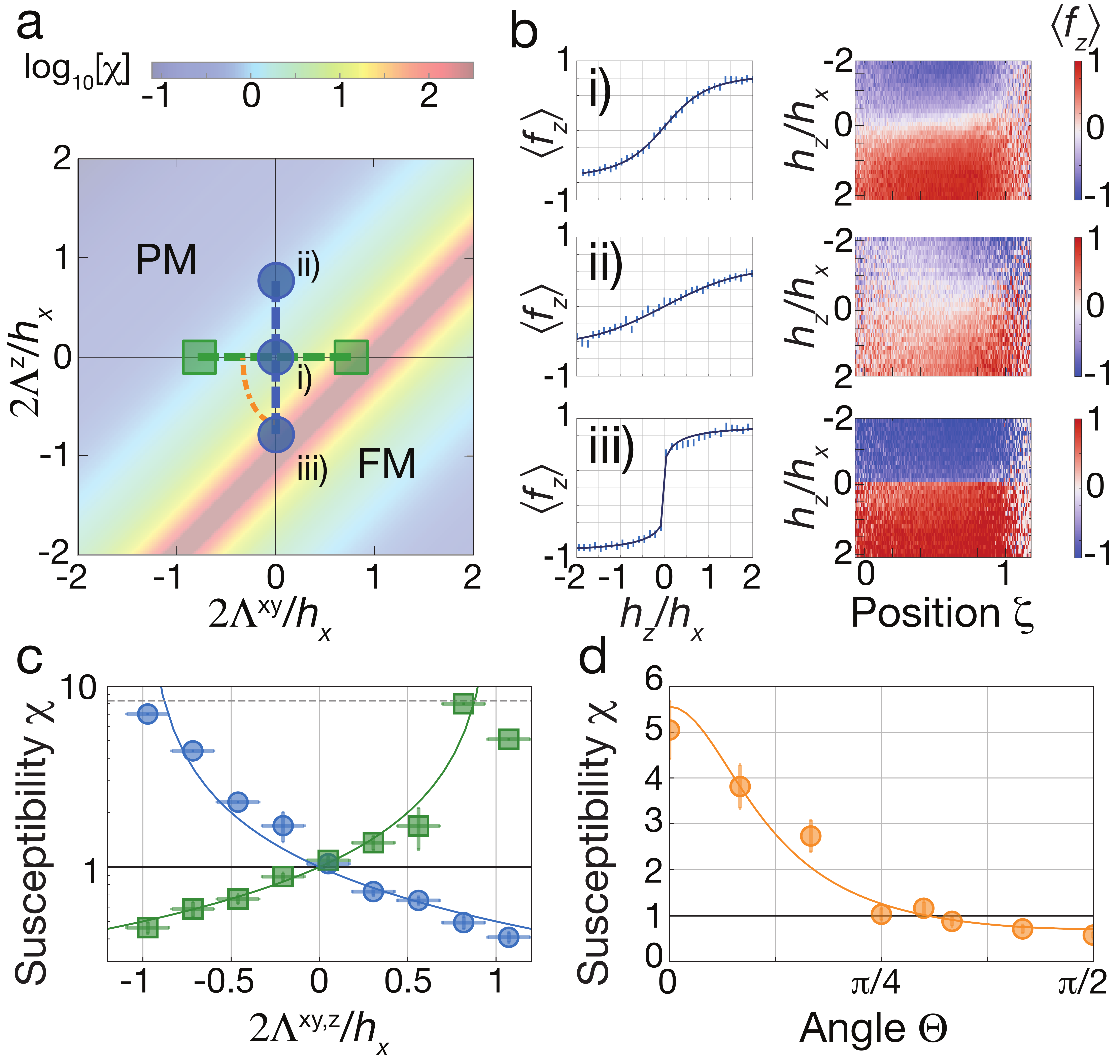}
\caption{Magnetic susceptibility of Ising and XY models. (a) Phase diagram of $H_\mathrm{XXZ}+h_x F_x$ for collective spin model in large-$\sF$ limit.  Color shows prediction for $\log[\chi]$, which diverges at transition between paramagnetic (PM) and ferromagnetic (FM) phases. Data for (b, c, d) and corresponding theory curves were taken along blue, green, and orange cuts. (b) Representative measurements of $\avg{f_z}$ vs $h_z$ at points i, ii, iii in phase diagram; $\chi(J^z)$ is extracted from spatially averaged data (left subplots with dark blue fit curves) \cite{SM}.  (c) Susceptibility $\chi$ vs $J^z$ (blue circles) and vs $J^{xy}$ (green squares). Dashed gray line indicates maximum detectable slope. (d) Susceptibility vs. $\th$ at fixed drive power. For non-interacting spins, $\chi = 1$  (black lines in c and d).}\label{fig:Figure3}
\end{figure}


The Hamiltonian can additionally be characterized by its low-energy states and their broken symmetries. To gain intuition for the phase diagram, we first consider the case where the cavity couples to a uniformly weighted collective spin $\vec{F}=\sum_i \vec{f}_i$. The total spin $F$ is then conserved, and the relation $\abs{\vec{F}}^2 - F_z^2 = F_x^2 + F_y^2$ reveals that any accessible Hamiltonian is equivalent to an Ising model with modified $J^{z}_{\text{eff}}\equiv J^z-J^{xy}$.  With an added transverse field $h_x\uvec{x}$, the system can undergo a phase transition from a paramagnet to an Ising ferromagnet with broken $\mathbb{Z}_2$ symmetry as a function of the effective Ising coupling $J^z_\mathrm{eff}$. Remarkably, in the collective spin picture we expect the ferromagnetic Ising phase also to exist in a system with antiferromagnetic XY interactions.

To test whether this prediction extends to our system with non-uniform interactions, we generate Hamiltonians of the generic form (with $\hbar = 1$)
\begin{equation}
\Htot \approx H_\mathrm{XXZ} + h_x F_x + h_z F_z
\end{equation}
by adding a global Raman coupling of Rabi frequency $h_x$ and detuning $h_z$.  Here, $H_\mathrm{XXZ}$ is the Hamiltonian of Eq. \ref{eq:H} with non-uniform couplings $\w_i$, whereas the Raman coupling and detuning are approximately uniform.  We prepare the paramagnetic ground state of $H_0 = h_x F_x + h_z F_z$ by adiabatically sweeping the detuning of the Raman beam from far off resonance to a final value $h_z$, at fixed Rabi frequency $h_x = 2\pi \times 2~\mathrm{kHz}$.  We then ramp on interactions $H_\mathrm{XXZ}$ over 5~ms to prepare a low-energy state of $\Htot$ and image the resulting magnetization.

Representative images of the magnetization vs the symmetry-breaking field $h_z$ are shown in Fig. \ref{fig:Figure3}b for different values of the Ising coupling, with $J^{xy}=0$.  In the absence of interactions ($J_z = 0$), the measured magnetization matches the prediction $\avg{f_z} = h_z/\sqrt{h_z^2 + h_x^2}$ [Fig. \ref{fig:Figure3}(b.i)].  Antiferromagnetic Ising interactions push the spins towards the equator of the Bloch sphere [Fig.~\ref{fig:Figure3}(b.ii)], thereby suppressing sensitivity to changes in $h_z$. By contrast, ferromagnetic interactions force the spin vector towards a pole determined by the sign of $h_z$ [Fig.~\ref{fig:Figure3}(b.iii)].
We summarize this behavior by plotting the magnetic susceptibility
\begin{equation}
\chi~\equiv~ \partial \cos \theta/\partial (h_z/h_x)|_{h_z=0}
\end{equation}
as a function of $J^z$ in Fig. \ref{fig:Figure3}(c) (blue circles).


Comparing the magnetic susceptibility in the Ising model with analogous measurements for pure XY interactions [green squares in Fig. \ref{fig:Figure3}(c)], we observe a striking symmetry under $J^z \leftrightarrow -J^{xy}$.  In both cases, the susceptibility rises sharply to the maximum value allowed by our resolution in $h_z$ (gray line) at a critical value of the collective interaction parameter $\Lambda^{z,xy} \equiv J^{z,xy}\abs{\bm{\sF}}$. 
The data are consistent with a classical model
$\chi~=~1/(2\Lambda^z_\mathrm{eff}/h_x + 1)$,
valid in the large-$\sF$ limit, which predicts a diverging susceptibility at the critical point $-2 \Lambda^z_\mathrm{eff} = h_x$
of the paramagnetic-to-ferromagnetic phase transition. The model also agrees with measurements obtained by varying $\Lambda^{z}_{\text{eff}}$ via the tuning angle $\th$ [Fig.~\ref{fig:Figure3}(d)], which traces out the orange cut in the phase diagram of Fig.~\ref{fig:Figure3}(a).

Notably, the collective-spin approximation describes the magnetic susceptibility well despite spatial inhomogeneities. In addition to the variation in atom-cavity coupling $\w(\zeta)$, a magnetic field gradient and inhomogeneous ac Stark shifts from the trapping lattice result in non-uniform $h_z$. The non-uniformity, evident in the magnetization of the non-interacting system [Fig. \ref{fig:Figure3}(b.i)], is suppressed by the ferromagnetic Ising interactions [Fig.~\ref{fig:Figure3}(b.iii)], which tend to align the spins.  However, even for antiferromagnetic interactions (both Ising and XY), the collective-spin model describes the data well, which we attribute to the spin-polarization of the initial state and the aligning effect of the field $h_x$.

In principle, XY interactions can protect the spin coherence even without the aligning field.  The XY Hamiltonian $H \propto -(\sF_x^2 + \sF_y^2) =  \sF_z^2 - \abs{\bm{\sF}}^2$ 
has an energy gap $\sim \abs{\bm{\sF}}$ between manifolds of different total spin that the analogous Ising model $H \propto \sF_z^2$ lacks. This gap is expected to protect against dephasing from inhomogeneous fields $H_\mathrm{inh} = \sum_i h_{i,z} f_{i,z}$ \cite{norcia2018cavity}.  To test this prediction, we directly compare the impact of inhomogeneous fields on systems with ferromagnetic XY and antiferromagnetic Ising interactions, which are equivalent except for the energy gap. 

\begin{figure}[htb]
\includegraphics[width=\columnwidth]{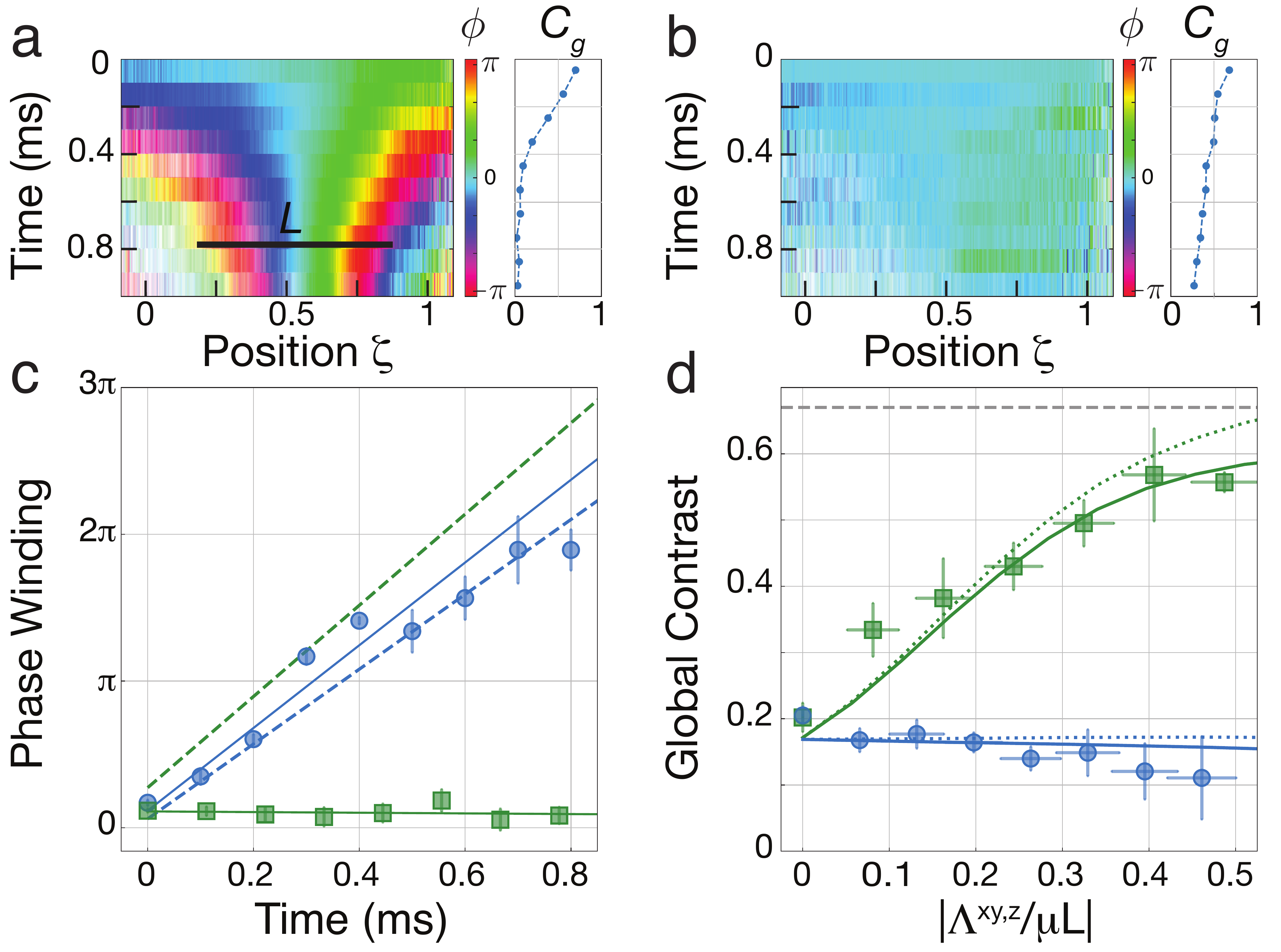}
\caption{Protection against dephasing. (a-b) Phase $\phi$ and global contrast $C_g$ at $\th = \pi/2$ for (a) no interactions and (b) XY interactions [$\Lambda^{xy}/\mu L = -0.43(4)$], in magnetic field gradient $\abs{\mu}=2\pi\times 2.1(1)~\mathrm{kHz}/z_R$. Opacity indicates length of transverse spin component. (c) Phase winding $\phi_L(t)$ for ferromagnetic XY interactions (squares, $\th=\pi/2$) or antiferromagnetic Ising interactions (circles, $\th=0$) of equal strength.  Dashed lines show $\phi_L$ in non-interacting systems at $\th = 0$ (blue) and $\th =\pi/2$ (green). (d) Global contrast $C_g$ vs collective interaction strength $\abs{\Lambda^{xy,z}}$ for XY (green squares) and Ising (blue circles) interactions, at $t = 0.5~$ms. Solid (dotted) curves show mean-field model with (without) free-space scattering. Dashed gray line indicates initial contrast $C_0$. }\label{fig:dephasing}
\end{figure}




To probe the robustness to inhomogeneous fields, we first adiabatically prepare a low-energy state of the Hamiltonian
$\Htot = H_\mathrm{XXZ} + h_x F_x + H_\mathrm{inh}$,
where $H_\mathrm{inh}$ consists of $\uvec{z}$-fields $h_{i,z} \lesssim h_x$.  We then quench off the aligning field $h_x$ and image the subsequent dynamics.  To calibrate the inhomogeneous field $H_\mathrm{inh}$, we first perform this quench without interactions and image the dephasing of the spins, as shown in Fig.~\ref{fig:dephasing}(a) for tuning angle $\th=\pi/2$.  We plot the total phase winding $\phi_L(t)$ across a length $L=1~$mm for $\th=\pi/2$ (green dashed line) and $\th=0$ (blue dashed line) in Fig.~\ref{fig:dephasing}(c).  For both angles, we observe similar magnitudes of the gradient $\mu = L^{-1} d\phi_L/dt$, as well as a small but nonzero initial phase winding $\phi_L(0)$ due to finite strength of the aligning field $h_x$ before the quench.  Whereas introducing Ising interactions has no effect on the dephasing (blue circles), XY interactions completely suppress the growth of phase winding (green squares). 


The onset of protected spin coherence is summarized in Fig.~\ref{fig:dephasing}(d), where we plot the global contrast $C_g \equiv \frac{1}{N}\sqrt{F_x^2 + F_y^2}$
vs interaction strength. We find an increase in contrast, indicating phasing of the spins, when the XY interaction strength becomes comparable to the gradient across the cloud, i.e., $\Lambda^{xy} \sim \mu L$. This condition can be understood in a mean-field picture by noting that the interactions produce an effective transverse field $\Lambda^{xy} \uvec{x}$, which must overcome the dephasing influence of differential $\uvec{z}$-fields of order $\mu L$~\cite{SM}. Equivalently, in a quantum mechanical description, interactions can protect against variations in $h_z$ that are small compared with the energy cost to flip a single spin, set by the gap of order $\Lambda^{xy}$ between sectors of different total spin~\cite{SM}.

Our intuitive understanding of the gap protection is confirmed by numerical simulations of the mean-field dynamics.  Both the data and simulations [solid curves in Fig. 4(d)] show the contrast increasing to a value $C_g \approx 0.6$, limited primarily by imperfect coherence $C_0 = 0.67(5)$ of the initial state.  A smaller effect included in the solid curves is spontaneous emission.  Comparing with an idealized model without spontaneous emission (dotted curves) shows that the interactions enhance coherence with minimal detriment from free-space scattering, thanks to the strong collective atom-light coupling.



In future work, spin-exchange interactions can be applied to maximize coherence in light-induced spin squeezing protocols \cite{leroux2010implementation,bohnet2014reduced,hosten2016measurement,hosten2016quantum,braverman2019near,wu2019retrieval}, operating either adiabatically or
via one-axis twisting dynamics \cite{hu2017vacuum,norcia2018cavity}.  
Notably, during one-axis twisting, the emergent many-body gap can protect the spin length without producing any additional dynamics, in contrast to an aligning transverse field that would cause unwanted rotations.
Gap protection could suppress even time-varying inhomogeneities arising from atomic motion, a benefit over spin echo pulses~\cite{schleier2010states}.  Extended to other platforms, gap protection can aid in preserving global spin coherence while using local (e.g., dipolar) interactions for entanglement generation \cite{rey2008many,cappellaro2009quantum,rudner2011generating, bouchoule2002spin,gil2014spin, kaubruegger2019variational, borish2020transverse}.

Much territory remains for further exploration of the nonlocal XXZ model. Perturbations of the antiferromagnetic Heisenberg model may yield chiral or valance-bond ground states \cite{mudry1989ground}, and the spin-1 structure may enrich the phase diagram  \cite{davis2019photon, luo2017deterministic,zhiqiang2017nonequilibrium, masson2017cavity}.  Adiabatic ramps could be used to prepare low-energy states of Ising models encoding combinatorial optimization problems~\cite{lucas2014ising}. 
Modifying the interaction graph by Floquet driving \cite{hung2016quantum} or local addressing can enable fast scrambling \cite{swingle2016measuring,bentsen2019treelike,marino2019cavity,kim2019low} or simulations of spin glasses \cite{gopalakrishnan2011frustration,strack2011dicke,buchhold2013dicke}.

\begin{acknowledgments}
E.~J.~D., A.~P., and E.~S.~C. contributed equally to this work. This work was supported by the DOE Office of Science, Office of High Energy Physics and Office of Basic Energy Sciences under grant No. DE-SC0019174.  A.~P. and E.~S.~C. acknowledge support from the NSF under grant No. PHY-1753021.  We additionally acknowledge support from the NSF Graduate Research Fellowship Program (E.~J.~D. and E.~S.~C.), from the Hertz Foundation (E.~J.~D.), and from the National Defense Science and Engineering Graduate Fellowship (A.~P.).
\end{acknowledgments}

\bibliography{references}

\begin{thebibliography}{68}%
\makeatletter
\providecommand \@ifxundefined [1]{%
 \@ifx{#1\undefined}
}%
\providecommand \@ifnum [1]{%
 \ifnum #1\expandafter \@firstoftwo
 \else \expandafter \@secondoftwo
 \fi
}%
\providecommand \@ifx [1]{%
 \ifx #1\expandafter \@firstoftwo
 \else \expandafter \@secondoftwo
 \fi
}%
\providecommand \natexlab [1]{#1}%
\providecommand \enquote  [1]{``#1''}%
\providecommand \bibnamefont  [1]{#1}%
\providecommand \bibfnamefont [1]{#1}%
\providecommand \citenamefont [1]{#1}%
\providecommand \href@noop [0]{\@secondoftwo}%
\providecommand \href [0]{\begingroup \@sanitize@url \@href}%
\providecommand \@href[1]{\@@startlink{#1}\@@href}%
\providecommand \@@href[1]{\endgroup#1\@@endlink}%
\providecommand \@sanitize@url [0]{\catcode `\\12\catcode `\$12\catcode
  `\&12\catcode `\#12\catcode `\^12\catcode `\_12\catcode `\%12\relax}%
\providecommand \@@startlink[1]{}%
\providecommand \@@endlink[0]{}%
\providecommand \url  [0]{\begingroup\@sanitize@url \@url }%
\providecommand \@url [1]{\endgroup\@href {#1}{\urlprefix }}%
\providecommand \urlprefix  [0]{URL }%
\providecommand \Eprint [0]{\href }%
\providecommand \doibase [0]{http://dx.doi.org/}%
\providecommand \selectlanguage [0]{\@gobble}%
\providecommand \bibinfo  [0]{\@secondoftwo}%
\providecommand \bibfield  [0]{\@secondoftwo}%
\providecommand \translation [1]{[#1]}%
\providecommand \BibitemOpen [0]{}%
\providecommand \bibitemStop [0]{}%
\providecommand \bibitemNoStop [0]{.\EOS\space}%
\providecommand \EOS [0]{\spacefactor3000\relax}%
\providecommand \BibitemShut  [1]{\csname bibitem#1\endcsname}%
\let\auto@bib@innerbib\@empty
\bibitem [{\citenamefont {Anderson}(1973)}]{anderson1973resonating}%
  \BibitemOpen
  \bibfield  {author} {\bibinfo {author} {\bibfnamefont {P.~W.}\ \bibnamefont
  {Anderson}},\ }\href@noop {} {\bibfield  {journal} {\bibinfo  {journal}
  {Materials Research Bulletin}\ }\textbf {\bibinfo {volume} {8}},\ \bibinfo
  {pages} {153} (\bibinfo {year} {1973})}\BibitemShut {NoStop}%
\bibitem [{\citenamefont {Kitaev}(2006)}]{kitaev2006anyons}%
  \BibitemOpen
  \bibfield  {author} {\bibinfo {author} {\bibfnamefont {A.}~\bibnamefont
  {Kitaev}},\ }\href@noop {} {\bibfield  {journal} {\bibinfo  {journal} {Annals
  of Physics}\ }\textbf {\bibinfo {volume} {321}},\ \bibinfo {pages} {2}
  (\bibinfo {year} {2006})}\BibitemShut {NoStop}%
\bibitem [{\citenamefont {Duan}\ \emph {et~al.}(2003)\citenamefont {Duan},
  \citenamefont {Demler},\ and\ \citenamefont {Lukin}}]{duan2003controlling}%
  \BibitemOpen
  \bibfield  {author} {\bibinfo {author} {\bibfnamefont {L.-M.}\ \bibnamefont
  {Duan}}, \bibinfo {author} {\bibfnamefont {E.}~\bibnamefont {Demler}}, \ and\
  \bibinfo {author} {\bibfnamefont {M.~D.}\ \bibnamefont {Lukin}},\ }\href@noop
  {} {\bibfield  {journal} {\bibinfo  {journal} {Physical Review Letters}\
  }\textbf {\bibinfo {volume} {91}},\ \bibinfo {pages} {090402} (\bibinfo
  {year} {2003})}\BibitemShut {NoStop}%
\bibitem [{\citenamefont {Barnett}\ \emph {et~al.}(2006)\citenamefont
  {Barnett}, \citenamefont {Petrov}, \citenamefont {Lukin},\ and\ \citenamefont
  {Demler}}]{barnett2006quantum}%
  \BibitemOpen
  \bibfield  {author} {\bibinfo {author} {\bibfnamefont {R.}~\bibnamefont
  {Barnett}}, \bibinfo {author} {\bibfnamefont {D.}~\bibnamefont {Petrov}},
  \bibinfo {author} {\bibfnamefont {M.}~\bibnamefont {Lukin}}, \ and\ \bibinfo
  {author} {\bibfnamefont {E.}~\bibnamefont {Demler}},\ }\href@noop {}
  {\bibfield  {journal} {\bibinfo  {journal} {Physical Review Letters}\
  }\textbf {\bibinfo {volume} {96}},\ \bibinfo {pages} {190401} (\bibinfo
  {year} {2006})}\BibitemShut {NoStop}%
\bibitem [{\citenamefont {Trotzky}\ \emph {et~al.}(2008)\citenamefont
  {Trotzky}, \citenamefont {Cheinet}, \citenamefont {F{\"o}lling},
  \citenamefont {Feld}, \citenamefont {Schnorrberger}, \citenamefont {Rey},
  \citenamefont {Polkovnikov}, \citenamefont {Demler}, \citenamefont {Lukin},\
  and\ \citenamefont {Bloch}}]{trotzky2008time}%
  \BibitemOpen
  \bibfield  {author} {\bibinfo {author} {\bibfnamefont {S.}~\bibnamefont
  {Trotzky}}, \bibinfo {author} {\bibfnamefont {P.}~\bibnamefont {Cheinet}},
  \bibinfo {author} {\bibfnamefont {S.}~\bibnamefont {F{\"o}lling}}, \bibinfo
  {author} {\bibfnamefont {M.}~\bibnamefont {Feld}}, \bibinfo {author}
  {\bibfnamefont {U.}~\bibnamefont {Schnorrberger}}, \bibinfo {author}
  {\bibfnamefont {A.~M.}\ \bibnamefont {Rey}}, \bibinfo {author} {\bibfnamefont
  {A.}~\bibnamefont {Polkovnikov}}, \bibinfo {author} {\bibfnamefont {E.~A.}\
  \bibnamefont {Demler}}, \bibinfo {author} {\bibfnamefont {M.~D.}\
  \bibnamefont {Lukin}}, \ and\ \bibinfo {author} {\bibfnamefont
  {I.}~\bibnamefont {Bloch}},\ }\href@noop {} {\bibfield  {journal} {\bibinfo
  {journal} {Science}\ }\textbf {\bibinfo {volume} {319}},\ \bibinfo {pages}
  {295} (\bibinfo {year} {2008})}\BibitemShut {NoStop}%
\bibitem [{\citenamefont {Simon}\ \emph {et~al.}(2011)\citenamefont {Simon},
  \citenamefont {Bakr}, \citenamefont {Ma}, \citenamefont {Tai}, \citenamefont
  {Preiss},\ and\ \citenamefont {Greiner}}]{simon2011quantum}%
  \BibitemOpen
  \bibfield  {author} {\bibinfo {author} {\bibfnamefont {J.}~\bibnamefont
  {Simon}}, \bibinfo {author} {\bibfnamefont {W.~S.}\ \bibnamefont {Bakr}},
  \bibinfo {author} {\bibfnamefont {R.}~\bibnamefont {Ma}}, \bibinfo {author}
  {\bibfnamefont {M.~E.}\ \bibnamefont {Tai}}, \bibinfo {author} {\bibfnamefont
  {P.~M.}\ \bibnamefont {Preiss}}, \ and\ \bibinfo {author} {\bibfnamefont
  {M.}~\bibnamefont {Greiner}},\ }\href@noop {} {\bibfield  {journal} {\bibinfo
   {journal} {Nature}\ }\textbf {\bibinfo {volume} {472}},\ \bibinfo {pages}
  {307} (\bibinfo {year} {2011})}\BibitemShut {NoStop}%
\bibitem [{\citenamefont {Browaeys}\ and\ \citenamefont
  {Lahaye}(2020)}]{browaeys2020many}%
  \BibitemOpen
  \bibfield  {author} {\bibinfo {author} {\bibfnamefont {A.}~\bibnamefont
  {Browaeys}}\ and\ \bibinfo {author} {\bibfnamefont {T.}~\bibnamefont
  {Lahaye}},\ }\href@noop {} {\bibfield  {journal} {\bibinfo  {journal} {Nature
  Physics}\ }\textbf {\bibinfo {volume} {16}},\ \bibinfo {pages} {132}
  (\bibinfo {year} {2020})}\BibitemShut {NoStop}%
\bibitem [{\citenamefont {Bernien}\ \emph {et~al.}(2017)\citenamefont
  {Bernien}, \citenamefont {Schwartz}, \citenamefont {Keesling}, \citenamefont
  {Levine}, \citenamefont {Omran}, \citenamefont {Pichler}, \citenamefont
  {Choi}, \citenamefont {Zibrov}, \citenamefont {Endres}, \citenamefont
  {Greiner} \emph {et~al.}}]{bernien2017probing}%
  \BibitemOpen
  \bibfield  {author} {\bibinfo {author} {\bibfnamefont {H.}~\bibnamefont
  {Bernien}}, \bibinfo {author} {\bibfnamefont {S.}~\bibnamefont {Schwartz}},
  \bibinfo {author} {\bibfnamefont {A.}~\bibnamefont {Keesling}}, \bibinfo
  {author} {\bibfnamefont {H.}~\bibnamefont {Levine}}, \bibinfo {author}
  {\bibfnamefont {A.}~\bibnamefont {Omran}}, \bibinfo {author} {\bibfnamefont
  {H.}~\bibnamefont {Pichler}}, \bibinfo {author} {\bibfnamefont
  {S.}~\bibnamefont {Choi}}, \bibinfo {author} {\bibfnamefont {A.~S.}\
  \bibnamefont {Zibrov}}, \bibinfo {author} {\bibfnamefont {M.}~\bibnamefont
  {Endres}}, \bibinfo {author} {\bibfnamefont {M.}~\bibnamefont {Greiner}},
  \emph {et~al.},\ }\href@noop {} {\bibfield  {journal} {\bibinfo  {journal}
  {Nature}\ }\textbf {\bibinfo {volume} {551}},\ \bibinfo {pages} {579}
  (\bibinfo {year} {2017})}\BibitemShut {NoStop}%
\bibitem [{\citenamefont {Zeiher}\ \emph {et~al.}(2017)\citenamefont {Zeiher},
  \citenamefont {Choi}, \citenamefont {Rubio-Abadal}, \citenamefont {Pohl},
  \citenamefont {van Bijnen}, \citenamefont {Bloch},\ and\ \citenamefont
  {Gross}}]{zeiher2017coherent}%
  \BibitemOpen
  \bibfield  {author} {\bibinfo {author} {\bibfnamefont {J.}~\bibnamefont
  {Zeiher}}, \bibinfo {author} {\bibfnamefont {J.-y.}\ \bibnamefont {Choi}},
  \bibinfo {author} {\bibfnamefont {A.}~\bibnamefont {Rubio-Abadal}}, \bibinfo
  {author} {\bibfnamefont {T.}~\bibnamefont {Pohl}}, \bibinfo {author}
  {\bibfnamefont {R.}~\bibnamefont {van Bijnen}}, \bibinfo {author}
  {\bibfnamefont {I.}~\bibnamefont {Bloch}}, \ and\ \bibinfo {author}
  {\bibfnamefont {C.}~\bibnamefont {Gross}},\ }\href@noop {} {\bibfield
  {journal} {\bibinfo  {journal} {Physical Review X}\ }\textbf {\bibinfo
  {volume} {7}},\ \bibinfo {pages} {041063} (\bibinfo {year}
  {2017})}\BibitemShut {NoStop}%
\bibitem [{\citenamefont {Guardado-Sanchez}\ \emph {et~al.}(2018)\citenamefont
  {Guardado-Sanchez}, \citenamefont {Brown}, \citenamefont {Mitra},
  \citenamefont {Devakul}, \citenamefont {Huse}, \citenamefont {Schau{\ss}},\
  and\ \citenamefont {Bakr}}]{guardado2018probing}%
  \BibitemOpen
  \bibfield  {author} {\bibinfo {author} {\bibfnamefont {E.}~\bibnamefont
  {Guardado-Sanchez}}, \bibinfo {author} {\bibfnamefont {P.~T.}\ \bibnamefont
  {Brown}}, \bibinfo {author} {\bibfnamefont {D.}~\bibnamefont {Mitra}},
  \bibinfo {author} {\bibfnamefont {T.}~\bibnamefont {Devakul}}, \bibinfo
  {author} {\bibfnamefont {D.~A.}\ \bibnamefont {Huse}}, \bibinfo {author}
  {\bibfnamefont {P.}~\bibnamefont {Schau{\ss}}}, \ and\ \bibinfo {author}
  {\bibfnamefont {W.~S.}\ \bibnamefont {Bakr}},\ }\href@noop {} {\bibfield
  {journal} {\bibinfo  {journal} {Physical Review X}\ }\textbf {\bibinfo
  {volume} {8}},\ \bibinfo {pages} {021069} (\bibinfo {year}
  {2018})}\BibitemShut {NoStop}%
\bibitem [{\citenamefont {Micheli}\ \emph {et~al.}(2006)\citenamefont
  {Micheli}, \citenamefont {Brennen},\ and\ \citenamefont
  {Zoller}}]{micheli2006toolbox}%
  \BibitemOpen
  \bibfield  {author} {\bibinfo {author} {\bibfnamefont {A.}~\bibnamefont
  {Micheli}}, \bibinfo {author} {\bibfnamefont {G.}~\bibnamefont {Brennen}}, \
  and\ \bibinfo {author} {\bibfnamefont {P.}~\bibnamefont {Zoller}},\
  }\href@noop {} {\bibfield  {journal} {\bibinfo  {journal} {Nature Physics}\
  }\textbf {\bibinfo {volume} {2}},\ \bibinfo {pages} {341} (\bibinfo {year}
  {2006})}\BibitemShut {NoStop}%
\bibitem [{\citenamefont {Glaetzle}\ \emph {et~al.}(2015)\citenamefont
  {Glaetzle}, \citenamefont {Dalmonte}, \citenamefont {Nath}, \citenamefont
  {Gross}, \citenamefont {Bloch},\ and\ \citenamefont
  {Zoller}}]{glaetzle2015designing}%
  \BibitemOpen
  \bibfield  {author} {\bibinfo {author} {\bibfnamefont {A.~W.}\ \bibnamefont
  {Glaetzle}}, \bibinfo {author} {\bibfnamefont {M.}~\bibnamefont {Dalmonte}},
  \bibinfo {author} {\bibfnamefont {R.}~\bibnamefont {Nath}}, \bibinfo {author}
  {\bibfnamefont {C.}~\bibnamefont {Gross}}, \bibinfo {author} {\bibfnamefont
  {I.}~\bibnamefont {Bloch}}, \ and\ \bibinfo {author} {\bibfnamefont
  {P.}~\bibnamefont {Zoller}},\ }\href {\doibase
  10.1103/PhysRevLett.114.173002} {\bibfield  {journal} {\bibinfo  {journal}
  {Phys. Rev. Lett.}\ }\textbf {\bibinfo {volume} {114}},\ \bibinfo {pages}
  {173002} (\bibinfo {year} {2015})}\BibitemShut {NoStop}%
\bibitem [{\citenamefont {Yao}\ \emph {et~al.}(2018)\citenamefont {Yao},
  \citenamefont {Zaletel}, \citenamefont {Stamper-Kurn},\ and\ \citenamefont
  {Vishwanath}}]{yao2018quantum}%
  \BibitemOpen
  \bibfield  {author} {\bibinfo {author} {\bibfnamefont {N.~Y.}\ \bibnamefont
  {Yao}}, \bibinfo {author} {\bibfnamefont {M.~P.}\ \bibnamefont {Zaletel}},
  \bibinfo {author} {\bibfnamefont {D.~M.}\ \bibnamefont {Stamper-Kurn}}, \
  and\ \bibinfo {author} {\bibfnamefont {A.}~\bibnamefont {Vishwanath}},\
  }\href@noop {} {\bibfield  {journal} {\bibinfo  {journal} {Nature Physics}\
  }\textbf {\bibinfo {volume} {14}},\ \bibinfo {pages} {405} (\bibinfo {year}
  {2018})}\BibitemShut {NoStop}%
\bibitem [{\citenamefont {Hung}\ \emph {et~al.}(2016)\citenamefont {Hung},
  \citenamefont {Gonz{\'a}lez-Tudela}, \citenamefont {Cirac},\ and\
  \citenamefont {Kimble}}]{hung2016quantum}%
  \BibitemOpen
  \bibfield  {author} {\bibinfo {author} {\bibfnamefont {C.-L.}\ \bibnamefont
  {Hung}}, \bibinfo {author} {\bibfnamefont {A.}~\bibnamefont
  {Gonz{\'a}lez-Tudela}}, \bibinfo {author} {\bibfnamefont {J.~I.}\
  \bibnamefont {Cirac}}, \ and\ \bibinfo {author} {\bibfnamefont
  {H.}~\bibnamefont {Kimble}},\ }\href@noop {} {\bibfield  {journal} {\bibinfo
  {journal} {Proceedings of the National Academy of Sciences}\ }\textbf
  {\bibinfo {volume} {113}},\ \bibinfo {pages} {E4946} (\bibinfo {year}
  {2016})}\BibitemShut {NoStop}%
\bibitem [{\citenamefont {Strack}\ and\ \citenamefont
  {Sachdev}(2011)}]{strack2011dicke}%
  \BibitemOpen
  \bibfield  {author} {\bibinfo {author} {\bibfnamefont {P.}~\bibnamefont
  {Strack}}\ and\ \bibinfo {author} {\bibfnamefont {S.}~\bibnamefont
  {Sachdev}},\ }\href@noop {} {\bibfield  {journal} {\bibinfo  {journal}
  {Physical Review Letters}\ }\textbf {\bibinfo {volume} {107}},\ \bibinfo
  {pages} {277202} (\bibinfo {year} {2011})}\BibitemShut {NoStop}%
\bibitem [{\citenamefont {Gopalakrishnan}\ \emph {et~al.}(2011)\citenamefont
  {Gopalakrishnan}, \citenamefont {Lev},\ and\ \citenamefont
  {Goldbart}}]{gopalakrishnan2011frustration}%
  \BibitemOpen
  \bibfield  {author} {\bibinfo {author} {\bibfnamefont {S.}~\bibnamefont
  {Gopalakrishnan}}, \bibinfo {author} {\bibfnamefont {B.~L.}\ \bibnamefont
  {Lev}}, \ and\ \bibinfo {author} {\bibfnamefont {P.~M.}\ \bibnamefont
  {Goldbart}},\ }\href@noop {} {\bibfield  {journal} {\bibinfo  {journal}
  {Physical Review Letters}\ }\textbf {\bibinfo {volume} {107}},\ \bibinfo
  {pages} {277201} (\bibinfo {year} {2011})}\BibitemShut {NoStop}%
\bibitem [{\citenamefont {Buchhold}\ \emph {et~al.}(2013)\citenamefont
  {Buchhold}, \citenamefont {Strack}, \citenamefont {Sachdev},\ and\
  \citenamefont {Diehl}}]{buchhold2013dicke}%
  \BibitemOpen
  \bibfield  {author} {\bibinfo {author} {\bibfnamefont {M.}~\bibnamefont
  {Buchhold}}, \bibinfo {author} {\bibfnamefont {P.}~\bibnamefont {Strack}},
  \bibinfo {author} {\bibfnamefont {S.}~\bibnamefont {Sachdev}}, \ and\
  \bibinfo {author} {\bibfnamefont {S.}~\bibnamefont {Diehl}},\ }\href@noop {}
  {\bibfield  {journal} {\bibinfo  {journal} {Physical Review A}\ }\textbf
  {\bibinfo {volume} {87}},\ \bibinfo {pages} {063622} (\bibinfo {year}
  {2013})}\BibitemShut {NoStop}%
\bibitem [{\citenamefont {Swingle}\ \emph {et~al.}(2016)\citenamefont
  {Swingle}, \citenamefont {Bentsen}, \citenamefont {Schleier-Smith},\ and\
  \citenamefont {Hayden}}]{swingle2016measuring}%
  \BibitemOpen
  \bibfield  {author} {\bibinfo {author} {\bibfnamefont {B.}~\bibnamefont
  {Swingle}}, \bibinfo {author} {\bibfnamefont {G.}~\bibnamefont {Bentsen}},
  \bibinfo {author} {\bibfnamefont {M.}~\bibnamefont {Schleier-Smith}}, \ and\
  \bibinfo {author} {\bibfnamefont {P.}~\bibnamefont {Hayden}},\ }\href@noop {}
  {\bibfield  {journal} {\bibinfo  {journal} {Physical Review A}\ }\textbf
  {\bibinfo {volume} {94}},\ \bibinfo {pages} {040302} (\bibinfo {year}
  {2016})}\BibitemShut {NoStop}%
\bibitem [{\citenamefont {Bentsen}\ \emph
  {et~al.}(2019{\natexlab{a}})\citenamefont {Bentsen}, \citenamefont
  {Hashizume}, \citenamefont {Buyskikh}, \citenamefont {Davis}, \citenamefont
  {Daley}, \citenamefont {Gubser},\ and\ \citenamefont
  {Schleier-Smith}}]{bentsen2019treelike}%
  \BibitemOpen
  \bibfield  {author} {\bibinfo {author} {\bibfnamefont {G.}~\bibnamefont
  {Bentsen}}, \bibinfo {author} {\bibfnamefont {T.}~\bibnamefont {Hashizume}},
  \bibinfo {author} {\bibfnamefont {A.~S.}\ \bibnamefont {Buyskikh}}, \bibinfo
  {author} {\bibfnamefont {E.~J.}\ \bibnamefont {Davis}}, \bibinfo {author}
  {\bibfnamefont {A.~J.}\ \bibnamefont {Daley}}, \bibinfo {author}
  {\bibfnamefont {S.~S.}\ \bibnamefont {Gubser}}, \ and\ \bibinfo {author}
  {\bibfnamefont {M.}~\bibnamefont {Schleier-Smith}},\ }\href {\doibase
  10.1103/PhysRevLett.123.130601} {\bibfield  {journal} {\bibinfo  {journal}
  {Phys. Rev. Lett.}\ }\textbf {\bibinfo {volume} {123}},\ \bibinfo {pages}
  {130601} (\bibinfo {year} {2019}{\natexlab{a}})}\BibitemShut {NoStop}%
\bibitem [{\citenamefont {Marino}\ and\ \citenamefont
  {Rey}(2019)}]{marino2019cavity}%
  \BibitemOpen
  \bibfield  {author} {\bibinfo {author} {\bibfnamefont {J.}~\bibnamefont
  {Marino}}\ and\ \bibinfo {author} {\bibfnamefont {A.}~\bibnamefont {Rey}},\
  }\href@noop {} {\bibfield  {journal} {\bibinfo  {journal} {Physical Review
  A}\ }\textbf {\bibinfo {volume} {99}},\ \bibinfo {pages} {051803} (\bibinfo
  {year} {2019})}\BibitemShut {NoStop}%
\bibitem [{\citenamefont {Mivehvar}\ \emph {et~al.}(2019)\citenamefont
  {Mivehvar}, \citenamefont {Ritsch},\ and\ \citenamefont
  {Piazza}}]{mivehvar2019cavity}%
  \BibitemOpen
  \bibfield  {author} {\bibinfo {author} {\bibfnamefont {F.}~\bibnamefont
  {Mivehvar}}, \bibinfo {author} {\bibfnamefont {H.}~\bibnamefont {Ritsch}}, \
  and\ \bibinfo {author} {\bibfnamefont {F.}~\bibnamefont {Piazza}},\ }\href
  {\doibase 10.1103/PhysRevLett.122.113603} {\bibfield  {journal} {\bibinfo
  {journal} {Phys. Rev. Lett.}\ }\textbf {\bibinfo {volume} {122}},\ \bibinfo
  {pages} {113603} (\bibinfo {year} {2019})}\BibitemShut {NoStop}%
\bibitem [{\citenamefont {Bentsen}\ \emph
  {et~al.}(2019{\natexlab{b}})\citenamefont {Bentsen}, \citenamefont
  {Potirniche}, \citenamefont {Bulchandani}, \citenamefont {Scaffidi},
  \citenamefont {Cao}, \citenamefont {Qi}, \citenamefont {Schleier-Smith},\
  and\ \citenamefont {Altman}}]{bentsen2019integrable}%
  \BibitemOpen
  \bibfield  {author} {\bibinfo {author} {\bibfnamefont {G.}~\bibnamefont
  {Bentsen}}, \bibinfo {author} {\bibfnamefont {I.-D.}\ \bibnamefont
  {Potirniche}}, \bibinfo {author} {\bibfnamefont {V.~B.}\ \bibnamefont
  {Bulchandani}}, \bibinfo {author} {\bibfnamefont {T.}~\bibnamefont
  {Scaffidi}}, \bibinfo {author} {\bibfnamefont {X.}~\bibnamefont {Cao}},
  \bibinfo {author} {\bibfnamefont {X.-L.}\ \bibnamefont {Qi}}, \bibinfo
  {author} {\bibfnamefont {M.}~\bibnamefont {Schleier-Smith}}, \ and\ \bibinfo
  {author} {\bibfnamefont {E.}~\bibnamefont {Altman}},\ }\href {\doibase
  10.1103/PhysRevX.9.041011} {\bibfield  {journal} {\bibinfo  {journal} {Phys.
  Rev. X}\ }\textbf {\bibinfo {volume} {9}},\ \bibinfo {pages} {041011}
  (\bibinfo {year} {2019}{\natexlab{b}})}\BibitemShut {NoStop}%
\bibitem [{\citenamefont {Kim}\ \emph {et~al.}(2019)\citenamefont {Kim},
  \citenamefont {Cao},\ and\ \citenamefont {Altman}}]{kim2019low}%
  \BibitemOpen
  \bibfield  {author} {\bibinfo {author} {\bibfnamefont {J.}~\bibnamefont
  {Kim}}, \bibinfo {author} {\bibfnamefont {X.}~\bibnamefont {Cao}}, \ and\
  \bibinfo {author} {\bibfnamefont {E.}~\bibnamefont {Altman}},\ }\href@noop {}
  {\bibfield  {journal} {\bibinfo  {journal} {arXiv:1910.10173
  [cond-mat.str-el]}\ } (\bibinfo {year} {2019})}\BibitemShut {NoStop}%
\bibitem [{\citenamefont {Defenu}\ \emph {et~al.}(2018)\citenamefont {Defenu},
  \citenamefont {Enss}, \citenamefont {Kastner},\ and\ \citenamefont
  {Morigi}}]{defenu2018dynamical}%
  \BibitemOpen
  \bibfield  {author} {\bibinfo {author} {\bibfnamefont {N.}~\bibnamefont
  {Defenu}}, \bibinfo {author} {\bibfnamefont {T.}~\bibnamefont {Enss}},
  \bibinfo {author} {\bibfnamefont {M.}~\bibnamefont {Kastner}}, \ and\
  \bibinfo {author} {\bibfnamefont {G.}~\bibnamefont {Morigi}},\ }\href@noop {}
  {\bibfield  {journal} {\bibinfo  {journal} {Physical Review Letters}\
  }\textbf {\bibinfo {volume} {121}},\ \bibinfo {pages} {240403} (\bibinfo
  {year} {2018})}\BibitemShut {NoStop}%
\bibitem [{\citenamefont {Colella}\ \emph {et~al.}(2018)\citenamefont
  {Colella}, \citenamefont {Citro}, \citenamefont {Barsanti}, \citenamefont
  {Rossini},\ and\ \citenamefont {Chiofalo}}]{colella2018quantum}%
  \BibitemOpen
  \bibfield  {author} {\bibinfo {author} {\bibfnamefont {E.}~\bibnamefont
  {Colella}}, \bibinfo {author} {\bibfnamefont {R.}~\bibnamefont {Citro}},
  \bibinfo {author} {\bibfnamefont {M.}~\bibnamefont {Barsanti}}, \bibinfo
  {author} {\bibfnamefont {D.}~\bibnamefont {Rossini}}, \ and\ \bibinfo
  {author} {\bibfnamefont {M.-L.}\ \bibnamefont {Chiofalo}},\ }\href@noop {}
  {\bibfield  {journal} {\bibinfo  {journal} {Physical Review B}\ }\textbf
  {\bibinfo {volume} {97}},\ \bibinfo {pages} {134502} (\bibinfo {year}
  {2018})}\BibitemShut {NoStop}%
\bibitem [{\citenamefont {Johnson}\ \emph {et~al.}(2011)\citenamefont
  {Johnson}, \citenamefont {Amin}, \citenamefont {Gildert}, \citenamefont
  {Lanting}, \citenamefont {Hamze}, \citenamefont {Dickson}, \citenamefont
  {Harris}, \citenamefont {Berkley}, \citenamefont {Johansson}, \citenamefont
  {Bunyk} \emph {et~al.}}]{johnson2011quantum}%
  \BibitemOpen
  \bibfield  {author} {\bibinfo {author} {\bibfnamefont {M.~W.}\ \bibnamefont
  {Johnson}}, \bibinfo {author} {\bibfnamefont {M.~H.}\ \bibnamefont {Amin}},
  \bibinfo {author} {\bibfnamefont {S.}~\bibnamefont {Gildert}}, \bibinfo
  {author} {\bibfnamefont {T.}~\bibnamefont {Lanting}}, \bibinfo {author}
  {\bibfnamefont {F.}~\bibnamefont {Hamze}}, \bibinfo {author} {\bibfnamefont
  {N.}~\bibnamefont {Dickson}}, \bibinfo {author} {\bibfnamefont
  {R.}~\bibnamefont {Harris}}, \bibinfo {author} {\bibfnamefont {A.~J.}\
  \bibnamefont {Berkley}}, \bibinfo {author} {\bibfnamefont {J.}~\bibnamefont
  {Johansson}}, \bibinfo {author} {\bibfnamefont {P.}~\bibnamefont {Bunyk}},
  \emph {et~al.},\ }\href@noop {} {\bibfield  {journal} {\bibinfo  {journal}
  {Nature}\ }\textbf {\bibinfo {volume} {473}},\ \bibinfo {pages} {194}
  (\bibinfo {year} {2011})}\BibitemShut {NoStop}%
\bibitem [{\citenamefont {Lucas}(2014)}]{lucas2014ising}%
  \BibitemOpen
  \bibfield  {author} {\bibinfo {author} {\bibfnamefont {A.}~\bibnamefont
  {Lucas}},\ }\href@noop {} {\bibfield  {journal} {\bibinfo  {journal}
  {Frontiers in Physics}\ }\textbf {\bibinfo {volume} {2}},\ \bibinfo {pages}
  {5} (\bibinfo {year} {2014})}\BibitemShut {NoStop}%
\bibitem [{\citenamefont {Morrison}\ and\ \citenamefont
  {Parkins}(2008)}]{morrison2008dynamical}%
  \BibitemOpen
  \bibfield  {author} {\bibinfo {author} {\bibfnamefont {S.}~\bibnamefont
  {Morrison}}\ and\ \bibinfo {author} {\bibfnamefont {A.}~\bibnamefont
  {Parkins}},\ }\href@noop {} {\bibfield  {journal} {\bibinfo  {journal}
  {Physical review letters}\ }\textbf {\bibinfo {volume} {100}},\ \bibinfo
  {pages} {040403} (\bibinfo {year} {2008})}\BibitemShut {NoStop}%
\bibitem [{\citenamefont {Chiacchio}\ and\ \citenamefont
  {Nunnenkamp}(2019)}]{chiacchio2019dissipation}%
  \BibitemOpen
  \bibfield  {author} {\bibinfo {author} {\bibfnamefont {E.~R.}\ \bibnamefont
  {Chiacchio}}\ and\ \bibinfo {author} {\bibfnamefont {A.}~\bibnamefont
  {Nunnenkamp}},\ }\href@noop {} {\bibfield  {journal} {\bibinfo  {journal}
  {Physical Review Letters}\ }\textbf {\bibinfo {volume} {122}},\ \bibinfo
  {pages} {193605} (\bibinfo {year} {2019})}\BibitemShut {NoStop}%
\bibitem [{\citenamefont {Muniz}\ \emph {et~al.}(2019)\citenamefont {Muniz},
  \citenamefont {Barberena}, \citenamefont {Lewis-Swan}, \citenamefont {Young},
  \citenamefont {Cline}, \citenamefont {Rey},\ and\ \citenamefont
  {Thompson}}]{muniz2019exploring}%
  \BibitemOpen
  \bibfield  {author} {\bibinfo {author} {\bibfnamefont {J.~A.}\ \bibnamefont
  {Muniz}}, \bibinfo {author} {\bibfnamefont {D.}~\bibnamefont {Barberena}},
  \bibinfo {author} {\bibfnamefont {R.~J.}\ \bibnamefont {Lewis-Swan}},
  \bibinfo {author} {\bibfnamefont {D.~J.}\ \bibnamefont {Young}}, \bibinfo
  {author} {\bibfnamefont {J.~R.}\ \bibnamefont {Cline}}, \bibinfo {author}
  {\bibfnamefont {A.~M.}\ \bibnamefont {Rey}}, \ and\ \bibinfo {author}
  {\bibfnamefont {J.~K.}\ \bibnamefont {Thompson}},\ }\href@noop {} {\bibfield
  {journal} {\bibinfo  {journal} {arXiv:1910.00439 [quant-ph]}\ } (\bibinfo
  {year} {2019})}\BibitemShut {NoStop}%
\bibitem [{\citenamefont {Leroux}\ \emph {et~al.}(2010)\citenamefont {Leroux},
  \citenamefont {Schleier-Smith},\ and\ \citenamefont
  {Vuleti\ifmmode~\acute{c}\else \'{c}\fi{}}}]{leroux2010implementation}%
  \BibitemOpen
  \bibfield  {author} {\bibinfo {author} {\bibfnamefont {I.~D.}\ \bibnamefont
  {Leroux}}, \bibinfo {author} {\bibfnamefont {M.~H.}\ \bibnamefont
  {Schleier-Smith}}, \ and\ \bibinfo {author} {\bibfnamefont {V.}~\bibnamefont
  {Vuleti\ifmmode~\acute{c}\else \'{c}\fi{}}},\ }\href {\doibase
  10.1103/PhysRevLett.104.073602} {\bibfield  {journal} {\bibinfo  {journal}
  {Phys. Rev. Lett.}\ }\textbf {\bibinfo {volume} {104}},\ \bibinfo {pages}
  {073602} (\bibinfo {year} {2010})}\BibitemShut {NoStop}%
\bibitem [{\citenamefont {Hosten}\ \emph
  {et~al.}(2016{\natexlab{a}})\citenamefont {Hosten}, \citenamefont
  {Krishnakumar}, \citenamefont {Engelsen},\ and\ \citenamefont
  {Kasevich}}]{hosten2016quantum}%
  \BibitemOpen
  \bibfield  {author} {\bibinfo {author} {\bibfnamefont {O.}~\bibnamefont
  {Hosten}}, \bibinfo {author} {\bibfnamefont {R.}~\bibnamefont
  {Krishnakumar}}, \bibinfo {author} {\bibfnamefont {N.~J.}\ \bibnamefont
  {Engelsen}}, \ and\ \bibinfo {author} {\bibfnamefont {M.~A.}\ \bibnamefont
  {Kasevich}},\ }\href@noop {} {\bibfield  {journal} {\bibinfo  {journal}
  {Science}\ }\textbf {\bibinfo {volume} {352}},\ \bibinfo {pages} {1552}
  (\bibinfo {year} {2016}{\natexlab{a}})}\BibitemShut {NoStop}%
\bibitem [{\citenamefont {Braverman}\ \emph {et~al.}(2019)\citenamefont
  {Braverman}, \citenamefont {Kawasaki}, \citenamefont {Pedrozo-Pe{\~n}afiel},
  \citenamefont {Colombo}, \citenamefont {Shu}, \citenamefont {Li},
  \citenamefont {Mendez}, \citenamefont {Yamoah}, \citenamefont {Salvi},
  \citenamefont {Akamatsu} \emph {et~al.}}]{braverman2019near}%
  \BibitemOpen
  \bibfield  {author} {\bibinfo {author} {\bibfnamefont {B.}~\bibnamefont
  {Braverman}}, \bibinfo {author} {\bibfnamefont {A.}~\bibnamefont {Kawasaki}},
  \bibinfo {author} {\bibfnamefont {E.}~\bibnamefont {Pedrozo-Pe{\~n}afiel}},
  \bibinfo {author} {\bibfnamefont {S.}~\bibnamefont {Colombo}}, \bibinfo
  {author} {\bibfnamefont {C.}~\bibnamefont {Shu}}, \bibinfo {author}
  {\bibfnamefont {Z.}~\bibnamefont {Li}}, \bibinfo {author} {\bibfnamefont
  {E.}~\bibnamefont {Mendez}}, \bibinfo {author} {\bibfnamefont
  {M.}~\bibnamefont {Yamoah}}, \bibinfo {author} {\bibfnamefont
  {L.}~\bibnamefont {Salvi}}, \bibinfo {author} {\bibfnamefont
  {D.}~\bibnamefont {Akamatsu}},  \emph {et~al.},\ }\href@noop {} {\bibfield
  {journal} {\bibinfo  {journal} {Physical Review Letters}\ }\textbf {\bibinfo
  {volume} {122}},\ \bibinfo {pages} {223203} (\bibinfo {year}
  {2019})}\BibitemShut {NoStop}%
\bibitem [{\citenamefont {Barontini}\ \emph {et~al.}(2015)\citenamefont
  {Barontini}, \citenamefont {Hohmann}, \citenamefont {Haas}, \citenamefont
  {Est{\`e}ve},\ and\ \citenamefont {Reichel}}]{barontini2015deterministic}%
  \BibitemOpen
  \bibfield  {author} {\bibinfo {author} {\bibfnamefont {G.}~\bibnamefont
  {Barontini}}, \bibinfo {author} {\bibfnamefont {L.}~\bibnamefont {Hohmann}},
  \bibinfo {author} {\bibfnamefont {F.}~\bibnamefont {Haas}}, \bibinfo {author}
  {\bibfnamefont {J.}~\bibnamefont {Est{\`e}ve}}, \ and\ \bibinfo {author}
  {\bibfnamefont {J.}~\bibnamefont {Reichel}},\ }\href@noop {} {\bibfield
  {journal} {\bibinfo  {journal} {Science}\ }\textbf {\bibinfo {volume}
  {349}},\ \bibinfo {pages} {1317} (\bibinfo {year} {2015})}\BibitemShut
  {NoStop}%
\bibitem [{\citenamefont {Davis}\ \emph {et~al.}(2016)\citenamefont {Davis},
  \citenamefont {Bentsen},\ and\ \citenamefont
  {Schleier-Smith}}]{davis2016approaching}%
  \BibitemOpen
  \bibfield  {author} {\bibinfo {author} {\bibfnamefont {E.}~\bibnamefont
  {Davis}}, \bibinfo {author} {\bibfnamefont {G.}~\bibnamefont {Bentsen}}, \
  and\ \bibinfo {author} {\bibfnamefont {M.}~\bibnamefont {Schleier-Smith}},\
  }\href@noop {} {\bibfield  {journal} {\bibinfo  {journal} {Physical Review
  Letters}\ }\textbf {\bibinfo {volume} {116}},\ \bibinfo {pages} {053601}
  (\bibinfo {year} {2016})}\BibitemShut {NoStop}%
\bibitem [{\citenamefont {Hu}\ \emph {et~al.}(2017)\citenamefont {Hu},
  \citenamefont {Chen}, \citenamefont {Vendeiro}, \citenamefont {Urvoy},
  \citenamefont {Braverman},\ and\ \citenamefont {Vuleti{\'c}}}]{hu2017vacuum}%
  \BibitemOpen
  \bibfield  {author} {\bibinfo {author} {\bibfnamefont {J.}~\bibnamefont
  {Hu}}, \bibinfo {author} {\bibfnamefont {W.}~\bibnamefont {Chen}}, \bibinfo
  {author} {\bibfnamefont {Z.}~\bibnamefont {Vendeiro}}, \bibinfo {author}
  {\bibfnamefont {A.}~\bibnamefont {Urvoy}}, \bibinfo {author} {\bibfnamefont
  {B.}~\bibnamefont {Braverman}}, \ and\ \bibinfo {author} {\bibfnamefont
  {V.}~\bibnamefont {Vuleti{\'c}}},\ }\href@noop {} {\bibfield  {journal}
  {\bibinfo  {journal} {Physical Review A}\ }\textbf {\bibinfo {volume} {96}},\
  \bibinfo {pages} {050301} (\bibinfo {year} {2017})}\BibitemShut {NoStop}%
\bibitem [{\citenamefont {Lewis-Swan}\ \emph {et~al.}(2018)\citenamefont
  {Lewis-Swan}, \citenamefont {Norcia}, \citenamefont {Cline}, \citenamefont
  {Thompson},\ and\ \citenamefont {Rey}}]{lewis2018robust}%
  \BibitemOpen
  \bibfield  {author} {\bibinfo {author} {\bibfnamefont {R.~J.}\ \bibnamefont
  {Lewis-Swan}}, \bibinfo {author} {\bibfnamefont {M.~A.}\ \bibnamefont
  {Norcia}}, \bibinfo {author} {\bibfnamefont {J.~R.}\ \bibnamefont {Cline}},
  \bibinfo {author} {\bibfnamefont {J.~K.}\ \bibnamefont {Thompson}}, \ and\
  \bibinfo {author} {\bibfnamefont {A.~M.}\ \bibnamefont {Rey}},\ }\href@noop
  {} {\bibfield  {journal} {\bibinfo  {journal} {Physical Review Letters}\
  }\textbf {\bibinfo {volume} {121}},\ \bibinfo {pages} {070403} (\bibinfo
  {year} {2018})}\BibitemShut {NoStop}%
\bibitem [{\citenamefont {Masson}\ \emph {et~al.}(2017)\citenamefont {Masson},
  \citenamefont {Barrett},\ and\ \citenamefont {Parkins}}]{masson2017cavity}%
  \BibitemOpen
  \bibfield  {author} {\bibinfo {author} {\bibfnamefont {S.~J.}\ \bibnamefont
  {Masson}}, \bibinfo {author} {\bibfnamefont {M.}~\bibnamefont {Barrett}}, \
  and\ \bibinfo {author} {\bibfnamefont {S.}~\bibnamefont {Parkins}},\
  }\href@noop {} {\bibfield  {journal} {\bibinfo  {journal} {Physical Review
  Letters}\ }\textbf {\bibinfo {volume} {119}},\ \bibinfo {pages} {213601}
  (\bibinfo {year} {2017})}\BibitemShut {NoStop}%
\bibitem [{\citenamefont {Vaidya}\ \emph {et~al.}(2018)\citenamefont {Vaidya},
  \citenamefont {Guo}, \citenamefont {Kroeze}, \citenamefont {Ballantine},
  \citenamefont {Koll{\'a}r}, \citenamefont {Keeling},\ and\ \citenamefont
  {Lev}}]{vaidya2018tunable}%
  \BibitemOpen
  \bibfield  {author} {\bibinfo {author} {\bibfnamefont {V.~D.}\ \bibnamefont
  {Vaidya}}, \bibinfo {author} {\bibfnamefont {Y.}~\bibnamefont {Guo}},
  \bibinfo {author} {\bibfnamefont {R.~M.}\ \bibnamefont {Kroeze}}, \bibinfo
  {author} {\bibfnamefont {K.~E.}\ \bibnamefont {Ballantine}}, \bibinfo
  {author} {\bibfnamefont {A.~J.}\ \bibnamefont {Koll{\'a}r}}, \bibinfo
  {author} {\bibfnamefont {J.}~\bibnamefont {Keeling}}, \ and\ \bibinfo
  {author} {\bibfnamefont {B.~L.}\ \bibnamefont {Lev}},\ }\href@noop {}
  {\bibfield  {journal} {\bibinfo  {journal} {Physical Review X}\ }\textbf
  {\bibinfo {volume} {8}},\ \bibinfo {pages} {011002} (\bibinfo {year}
  {2018})}\BibitemShut {NoStop}%
\bibitem [{\citenamefont {Norcia}\ \emph {et~al.}(2018)\citenamefont {Norcia},
  \citenamefont {Lewis-Swan}, \citenamefont {Cline}, \citenamefont {Zhu},
  \citenamefont {Rey},\ and\ \citenamefont {Thompson}}]{norcia2018cavity}%
  \BibitemOpen
  \bibfield  {author} {\bibinfo {author} {\bibfnamefont {M.~A.}\ \bibnamefont
  {Norcia}}, \bibinfo {author} {\bibfnamefont {R.~J.}\ \bibnamefont
  {Lewis-Swan}}, \bibinfo {author} {\bibfnamefont {J.~R.}\ \bibnamefont
  {Cline}}, \bibinfo {author} {\bibfnamefont {B.}~\bibnamefont {Zhu}}, \bibinfo
  {author} {\bibfnamefont {A.~M.}\ \bibnamefont {Rey}}, \ and\ \bibinfo
  {author} {\bibfnamefont {J.~K.}\ \bibnamefont {Thompson}},\ }\href@noop {}
  {\bibfield  {journal} {\bibinfo  {journal} {Science}\ }\textbf {\bibinfo
  {volume} {361}},\ \bibinfo {pages} {259} (\bibinfo {year}
  {2018})}\BibitemShut {NoStop}%
\bibitem [{\citenamefont {Landini}\ \emph {et~al.}(2018)\citenamefont
  {Landini}, \citenamefont {Dogra}, \citenamefont {Kr{\"o}ger}, \citenamefont
  {Hruby}, \citenamefont {Donner},\ and\ \citenamefont
  {Esslinger}}]{landini2018formation}%
  \BibitemOpen
  \bibfield  {author} {\bibinfo {author} {\bibfnamefont {M.}~\bibnamefont
  {Landini}}, \bibinfo {author} {\bibfnamefont {N.}~\bibnamefont {Dogra}},
  \bibinfo {author} {\bibfnamefont {K.}~\bibnamefont {Kr{\"o}ger}}, \bibinfo
  {author} {\bibfnamefont {L.}~\bibnamefont {Hruby}}, \bibinfo {author}
  {\bibfnamefont {T.}~\bibnamefont {Donner}}, \ and\ \bibinfo {author}
  {\bibfnamefont {T.}~\bibnamefont {Esslinger}},\ }\href@noop {} {\bibfield
  {journal} {\bibinfo  {journal} {Physical Review Letters}\ }\textbf {\bibinfo
  {volume} {120}},\ \bibinfo {pages} {223602} (\bibinfo {year}
  {2018})}\BibitemShut {NoStop}%
\bibitem [{\citenamefont {Davis}\ \emph {et~al.}(2019)\citenamefont {Davis},
  \citenamefont {Bentsen}, \citenamefont {Homeier}, \citenamefont {Li},\ and\
  \citenamefont {Schleier-Smith}}]{davis2019photon}%
  \BibitemOpen
  \bibfield  {author} {\bibinfo {author} {\bibfnamefont {E.~J.}\ \bibnamefont
  {Davis}}, \bibinfo {author} {\bibfnamefont {G.}~\bibnamefont {Bentsen}},
  \bibinfo {author} {\bibfnamefont {L.}~\bibnamefont {Homeier}}, \bibinfo
  {author} {\bibfnamefont {T.}~\bibnamefont {Li}}, \ and\ \bibinfo {author}
  {\bibfnamefont {M.~H.}\ \bibnamefont {Schleier-Smith}},\ }\href {\doibase
  10.1103/PhysRevLett.122.010405} {\bibfield  {journal} {\bibinfo  {journal}
  {Phys. Rev. Lett.}\ }\textbf {\bibinfo {volume} {122}},\ \bibinfo {pages}
  {010405} (\bibinfo {year} {2019})},\ \bibinfo {note} {also see Supplemental
  Material.}\BibitemShut {Stop}%
\bibitem [{\citenamefont {Spethmann}\ \emph {et~al.}(2016)\citenamefont
  {Spethmann}, \citenamefont {Kohler}, \citenamefont {Schreppler},
  \citenamefont {Buchmann},\ and\ \citenamefont
  {Stamper-Kurn}}]{spethmann2016cavity}%
  \BibitemOpen
  \bibfield  {author} {\bibinfo {author} {\bibfnamefont {N.}~\bibnamefont
  {Spethmann}}, \bibinfo {author} {\bibfnamefont {J.}~\bibnamefont {Kohler}},
  \bibinfo {author} {\bibfnamefont {S.}~\bibnamefont {Schreppler}}, \bibinfo
  {author} {\bibfnamefont {L.}~\bibnamefont {Buchmann}}, \ and\ \bibinfo
  {author} {\bibfnamefont {D.~M.}\ \bibnamefont {Stamper-Kurn}},\ }\href@noop
  {} {\bibfield  {journal} {\bibinfo  {journal} {Nature Physics}\ }\textbf
  {\bibinfo {volume} {12}},\ \bibinfo {pages} {27} (\bibinfo {year}
  {2016})}\BibitemShut {NoStop}%
\bibitem [{\citenamefont {Georges}\ \emph {et~al.}(2018)\citenamefont
  {Georges}, \citenamefont {Cosme}, \citenamefont {Mathey},\ and\ \citenamefont
  {Hemmerich}}]{georges2018light}%
  \BibitemOpen
  \bibfield  {author} {\bibinfo {author} {\bibfnamefont {C.}~\bibnamefont
  {Georges}}, \bibinfo {author} {\bibfnamefont {J.~G.}\ \bibnamefont {Cosme}},
  \bibinfo {author} {\bibfnamefont {L.}~\bibnamefont {Mathey}}, \ and\ \bibinfo
  {author} {\bibfnamefont {A.}~\bibnamefont {Hemmerich}},\ }\href@noop {}
  {\bibfield  {journal} {\bibinfo  {journal} {Physical Review Letters}\
  }\textbf {\bibinfo {volume} {121}},\ \bibinfo {pages} {220405} (\bibinfo
  {year} {2018})}\BibitemShut {NoStop}%
\bibitem [{\citenamefont {Kuzmich}\ and\ \citenamefont
  {Kennedy}(2004)}]{kuzmich2004nonsymmetric}%
  \BibitemOpen
  \bibfield  {author} {\bibinfo {author} {\bibfnamefont {A.}~\bibnamefont
  {Kuzmich}}\ and\ \bibinfo {author} {\bibfnamefont {T.}~\bibnamefont
  {Kennedy}},\ }\href@noop {} {\bibfield  {journal} {\bibinfo  {journal}
  {Physical Review Letters}\ }\textbf {\bibinfo {volume} {92}},\ \bibinfo
  {pages} {030407} (\bibinfo {year} {2004})}\BibitemShut {NoStop}%
\bibitem [{\citenamefont {Hu}\ \emph {et~al.}(2015)\citenamefont {Hu},
  \citenamefont {Chen}, \citenamefont {Vendeiro}, \citenamefont {Zhang},\ and\
  \citenamefont {Vuleti{\'c}}}]{hu2015entangled}%
  \BibitemOpen
  \bibfield  {author} {\bibinfo {author} {\bibfnamefont {J.}~\bibnamefont
  {Hu}}, \bibinfo {author} {\bibfnamefont {W.}~\bibnamefont {Chen}}, \bibinfo
  {author} {\bibfnamefont {Z.}~\bibnamefont {Vendeiro}}, \bibinfo {author}
  {\bibfnamefont {H.}~\bibnamefont {Zhang}}, \ and\ \bibinfo {author}
  {\bibfnamefont {V.}~\bibnamefont {Vuleti{\'c}}},\ }\href@noop {} {\bibfield
  {journal} {\bibinfo  {journal} {Physical Review A}\ }\textbf {\bibinfo
  {volume} {92}},\ \bibinfo {pages} {063816} (\bibinfo {year}
  {2015})}\BibitemShut {NoStop}%
\bibitem [{\citenamefont {Dellantonio}\ \emph {et~al.}(2017)\citenamefont
  {Dellantonio}, \citenamefont {Das}, \citenamefont {Appel},\ and\
  \citenamefont {S\o{}rensen}}]{dellantonio2017multiparticle}%
  \BibitemOpen
  \bibfield  {author} {\bibinfo {author} {\bibfnamefont {L.}~\bibnamefont
  {Dellantonio}}, \bibinfo {author} {\bibfnamefont {S.}~\bibnamefont {Das}},
  \bibinfo {author} {\bibfnamefont {J.}~\bibnamefont {Appel}}, \ and\ \bibinfo
  {author} {\bibfnamefont {A.~S.}\ \bibnamefont {S\o{}rensen}},\ }\href
  {\doibase 10.1103/PhysRevA.95.040301} {\bibfield  {journal} {\bibinfo
  {journal} {Phys. Rev. A}\ }\textbf {\bibinfo {volume} {95}},\ \bibinfo
  {pages} {040301} (\bibinfo {year} {2017})}\BibitemShut {NoStop}%
\bibitem [{\citenamefont {Wu}\ \emph {et~al.}(2019)\citenamefont {Wu},
  \citenamefont {Krishnakumar}, \citenamefont {Mart{\'\i}nez-Rinc{\'o}n},
  \citenamefont {Malia}, \citenamefont {Hosten},\ and\ \citenamefont
  {Kasevich}}]{wu2019retrieval}%
  \BibitemOpen
  \bibfield  {author} {\bibinfo {author} {\bibfnamefont {Y.}~\bibnamefont
  {Wu}}, \bibinfo {author} {\bibfnamefont {R.}~\bibnamefont {Krishnakumar}},
  \bibinfo {author} {\bibfnamefont {J.}~\bibnamefont
  {Mart{\'\i}nez-Rinc{\'o}n}}, \bibinfo {author} {\bibfnamefont {B.~K.}\
  \bibnamefont {Malia}}, \bibinfo {author} {\bibfnamefont {O.}~\bibnamefont
  {Hosten}}, \ and\ \bibinfo {author} {\bibfnamefont {M.~A.}\ \bibnamefont
  {Kasevich}},\ }\href@noop {} {\bibfield  {journal} {\bibinfo  {journal}
  {arXiv:1912.08334 [quant-ph]}\ } (\bibinfo {year} {2019})}\BibitemShut
  {NoStop}%
\bibitem [{\citenamefont {Deutsch}\ \emph {et~al.}(2010)\citenamefont
  {Deutsch}, \citenamefont {Ramirez-Martinez}, \citenamefont {Lacro{\^u}te},
  \citenamefont {Reinhard}, \citenamefont {Schneider}, \citenamefont {Fuchs},
  \citenamefont {Pi{\'e}chon}, \citenamefont {Lalo{\"e}}, \citenamefont
  {Reichel},\ and\ \citenamefont {Rosenbusch}}]{deutsch2010spin}%
  \BibitemOpen
  \bibfield  {author} {\bibinfo {author} {\bibfnamefont {C.}~\bibnamefont
  {Deutsch}}, \bibinfo {author} {\bibfnamefont {F.}~\bibnamefont
  {Ramirez-Martinez}}, \bibinfo {author} {\bibfnamefont {C.}~\bibnamefont
  {Lacro{\^u}te}}, \bibinfo {author} {\bibfnamefont {F.}~\bibnamefont
  {Reinhard}}, \bibinfo {author} {\bibfnamefont {T.}~\bibnamefont {Schneider}},
  \bibinfo {author} {\bibfnamefont {J.-N.}\ \bibnamefont {Fuchs}}, \bibinfo
  {author} {\bibfnamefont {F.}~\bibnamefont {Pi{\'e}chon}}, \bibinfo {author}
  {\bibfnamefont {F.}~\bibnamefont {Lalo{\"e}}}, \bibinfo {author}
  {\bibfnamefont {J.}~\bibnamefont {Reichel}}, \ and\ \bibinfo {author}
  {\bibfnamefont {P.}~\bibnamefont {Rosenbusch}},\ }\href@noop {} {\bibfield
  {journal} {\bibinfo  {journal} {Physical Review Letters}\ }\textbf {\bibinfo
  {volume} {105}},\ \bibinfo {pages} {020401} (\bibinfo {year}
  {2010})}\BibitemShut {NoStop}%
\bibitem [{\citenamefont {B{\"u}ning}\ \emph {et~al.}(2011)\citenamefont
  {B{\"u}ning}, \citenamefont {Will}, \citenamefont {Ertmer}, \citenamefont
  {Rasel}, \citenamefont {Arlt}, \citenamefont {Klempt}, \citenamefont
  {Ramirez-Martinez}, \citenamefont {Pi{\'e}chon},\ and\ \citenamefont
  {Rosenbusch}}]{buening2011extended}%
  \BibitemOpen
  \bibfield  {author} {\bibinfo {author} {\bibfnamefont {G.~K.}\ \bibnamefont
  {B{\"u}ning}}, \bibinfo {author} {\bibfnamefont {J.}~\bibnamefont {Will}},
  \bibinfo {author} {\bibfnamefont {W.}~\bibnamefont {Ertmer}}, \bibinfo
  {author} {\bibfnamefont {E.}~\bibnamefont {Rasel}}, \bibinfo {author}
  {\bibfnamefont {J.}~\bibnamefont {Arlt}}, \bibinfo {author} {\bibfnamefont
  {C.}~\bibnamefont {Klempt}}, \bibinfo {author} {\bibfnamefont
  {F.}~\bibnamefont {Ramirez-Martinez}}, \bibinfo {author} {\bibfnamefont
  {F.}~\bibnamefont {Pi{\'e}chon}}, \ and\ \bibinfo {author} {\bibfnamefont
  {P.}~\bibnamefont {Rosenbusch}},\ }\href@noop {} {\bibfield  {journal}
  {\bibinfo  {journal} {Physical Review Letters}\ }\textbf {\bibinfo {volume}
  {106}},\ \bibinfo {pages} {240801} (\bibinfo {year} {2011})}\BibitemShut
  {NoStop}%
\bibitem [{\citenamefont {Choi}\ \emph {et~al.}(2017)\citenamefont {Choi},
  \citenamefont {Choi}, \citenamefont {Landig}, \citenamefont {Kucsko},
  \citenamefont {Zhou}, \citenamefont {Isoya}, \citenamefont {Jelezko},
  \citenamefont {Onoda}, \citenamefont {Sumiya}, \citenamefont {Khemani} \emph
  {et~al.}}]{choi2017observation}%
  \BibitemOpen
  \bibfield  {author} {\bibinfo {author} {\bibfnamefont {S.}~\bibnamefont
  {Choi}}, \bibinfo {author} {\bibfnamefont {J.}~\bibnamefont {Choi}}, \bibinfo
  {author} {\bibfnamefont {R.}~\bibnamefont {Landig}}, \bibinfo {author}
  {\bibfnamefont {G.}~\bibnamefont {Kucsko}}, \bibinfo {author} {\bibfnamefont
  {H.}~\bibnamefont {Zhou}}, \bibinfo {author} {\bibfnamefont {J.}~\bibnamefont
  {Isoya}}, \bibinfo {author} {\bibfnamefont {F.}~\bibnamefont {Jelezko}},
  \bibinfo {author} {\bibfnamefont {S.}~\bibnamefont {Onoda}}, \bibinfo
  {author} {\bibfnamefont {H.}~\bibnamefont {Sumiya}}, \bibinfo {author}
  {\bibfnamefont {V.}~\bibnamefont {Khemani}},  \emph {et~al.},\ }\href@noop {}
  {\bibfield  {journal} {\bibinfo  {journal} {Nature}\ }\textbf {\bibinfo
  {volume} {543}},\ \bibinfo {pages} {221} (\bibinfo {year}
  {2017})}\BibitemShut {NoStop}%
\bibitem [{\citenamefont {Zhang}\ \emph {et~al.}(2017)\citenamefont {Zhang},
  \citenamefont {Hess}, \citenamefont {Kyprianidis}, \citenamefont {Becker},
  \citenamefont {Lee}, \citenamefont {Smith}, \citenamefont {Pagano},
  \citenamefont {Potirniche}, \citenamefont {Potter}, \citenamefont
  {Vishwanath} \emph {et~al.}}]{zhang2017observation}%
  \BibitemOpen
  \bibfield  {author} {\bibinfo {author} {\bibfnamefont {J.}~\bibnamefont
  {Zhang}}, \bibinfo {author} {\bibfnamefont {P.}~\bibnamefont {Hess}},
  \bibinfo {author} {\bibfnamefont {A.}~\bibnamefont {Kyprianidis}}, \bibinfo
  {author} {\bibfnamefont {P.}~\bibnamefont {Becker}}, \bibinfo {author}
  {\bibfnamefont {A.}~\bibnamefont {Lee}}, \bibinfo {author} {\bibfnamefont
  {J.}~\bibnamefont {Smith}}, \bibinfo {author} {\bibfnamefont
  {G.}~\bibnamefont {Pagano}}, \bibinfo {author} {\bibfnamefont {I.-D.}\
  \bibnamefont {Potirniche}}, \bibinfo {author} {\bibfnamefont {A.~C.}\
  \bibnamefont {Potter}}, \bibinfo {author} {\bibfnamefont {A.}~\bibnamefont
  {Vishwanath}},  \emph {et~al.},\ }\href@noop {} {\bibfield  {journal}
  {\bibinfo  {journal} {Nature}\ }\textbf {\bibinfo {volume} {543}},\ \bibinfo
  {pages} {217} (\bibinfo {year} {2017})}\BibitemShut {NoStop}%
\bibitem [{\citenamefont {Hammerer}\ \emph {et~al.}(2010)\citenamefont
  {Hammerer}, \citenamefont {S{\o}rensen},\ and\ \citenamefont
  {Polzik}}]{hammerer2010quantum}%
  \BibitemOpen
  \bibfield  {author} {\bibinfo {author} {\bibfnamefont {K.}~\bibnamefont
  {Hammerer}}, \bibinfo {author} {\bibfnamefont {A.~S.}\ \bibnamefont
  {S{\o}rensen}}, \ and\ \bibinfo {author} {\bibfnamefont {E.~S.}\ \bibnamefont
  {Polzik}},\ }\href@noop {} {\bibfield  {journal} {\bibinfo  {journal}
  {Reviews of Modern Physics}\ }\textbf {\bibinfo {volume} {82}},\ \bibinfo
  {pages} {1041} (\bibinfo {year} {2010})}\BibitemShut {NoStop}%
\bibitem [{\citenamefont {Kohler}\ \emph {et~al.}(2017)\citenamefont {Kohler},
  \citenamefont {Spethmann}, \citenamefont {Schreppler},\ and\ \citenamefont
  {Stamper-Kurn}}]{kohler2017cavity}%
  \BibitemOpen
  \bibfield  {author} {\bibinfo {author} {\bibfnamefont {J.}~\bibnamefont
  {Kohler}}, \bibinfo {author} {\bibfnamefont {N.}~\bibnamefont {Spethmann}},
  \bibinfo {author} {\bibfnamefont {S.}~\bibnamefont {Schreppler}}, \ and\
  \bibinfo {author} {\bibfnamefont {D.~M.}\ \bibnamefont {Stamper-Kurn}},\
  }\href@noop {} {\bibfield  {journal} {\bibinfo  {journal} {Physical Review
  Letters}\ }\textbf {\bibinfo {volume} {118}},\ \bibinfo {pages} {063604}
  (\bibinfo {year} {2017})}\BibitemShut {NoStop}%
\bibitem [{SM()}]{SM}%
  \BibitemOpen
  \href@noop {} {}\bibinfo {note} {See Supplemental Material at [URL will be
  inserted by publisher] for supporting derivations, including Refs.
  XXX.}\BibitemShut {Stop}%
\bibitem [{\citenamefont {Bohnet}\ \emph {et~al.}(2014)\citenamefont {Bohnet},
  \citenamefont {Cox}, \citenamefont {Norcia}, \citenamefont {Weiner},
  \citenamefont {Chen},\ and\ \citenamefont {Thompson}}]{bohnet2014reduced}%
  \BibitemOpen
  \bibfield  {author} {\bibinfo {author} {\bibfnamefont {J.~G.}\ \bibnamefont
  {Bohnet}}, \bibinfo {author} {\bibfnamefont {K.~C.}\ \bibnamefont {Cox}},
  \bibinfo {author} {\bibfnamefont {M.~A.}\ \bibnamefont {Norcia}}, \bibinfo
  {author} {\bibfnamefont {J.~M.}\ \bibnamefont {Weiner}}, \bibinfo {author}
  {\bibfnamefont {Z.}~\bibnamefont {Chen}}, \ and\ \bibinfo {author}
  {\bibfnamefont {J.~K.}\ \bibnamefont {Thompson}},\ }\href@noop {} {\bibfield
  {journal} {\bibinfo  {journal} {Nature Photonics}\ }\textbf {\bibinfo
  {volume} {8}},\ \bibinfo {pages} {731} (\bibinfo {year} {2014})}\BibitemShut
  {NoStop}%
\bibitem [{\citenamefont {Hosten}\ \emph
  {et~al.}(2016{\natexlab{b}})\citenamefont {Hosten}, \citenamefont {Engelsen},
  \citenamefont {Krishnakumar},\ and\ \citenamefont
  {Kasevich}}]{hosten2016measurement}%
  \BibitemOpen
  \bibfield  {author} {\bibinfo {author} {\bibfnamefont {O.}~\bibnamefont
  {Hosten}}, \bibinfo {author} {\bibfnamefont {N.~J.}\ \bibnamefont
  {Engelsen}}, \bibinfo {author} {\bibfnamefont {R.}~\bibnamefont
  {Krishnakumar}}, \ and\ \bibinfo {author} {\bibfnamefont {M.~A.}\
  \bibnamefont {Kasevich}},\ }\href@noop {} {\bibfield  {journal} {\bibinfo
  {journal} {Nature}\ }\textbf {\bibinfo {volume} {529}},\ \bibinfo {pages}
  {505} (\bibinfo {year} {2016}{\natexlab{b}})}\BibitemShut {NoStop}%
\bibitem [{\citenamefont {Schleier-Smith}\ \emph {et~al.}(2010)\citenamefont
  {Schleier-Smith}, \citenamefont {Leroux},\ and\ \citenamefont
  {Vuleti{\'c}}}]{schleier2010states}%
  \BibitemOpen
  \bibfield  {author} {\bibinfo {author} {\bibfnamefont {M.~H.}\ \bibnamefont
  {Schleier-Smith}}, \bibinfo {author} {\bibfnamefont {I.~D.}\ \bibnamefont
  {Leroux}}, \ and\ \bibinfo {author} {\bibfnamefont {V.}~\bibnamefont
  {Vuleti{\'c}}},\ }\href@noop {} {\bibfield  {journal} {\bibinfo  {journal}
  {Physical Review Letters}\ }\textbf {\bibinfo {volume} {104}},\ \bibinfo
  {pages} {073604} (\bibinfo {year} {2010})}\BibitemShut {NoStop}%
\bibitem [{\citenamefont {Rey}\ \emph {et~al.}(2008)\citenamefont {Rey},
  \citenamefont {Jiang}, \citenamefont {Fleischhauer}, \citenamefont {Demler},\
  and\ \citenamefont {Lukin}}]{rey2008many}%
  \BibitemOpen
  \bibfield  {author} {\bibinfo {author} {\bibfnamefont {A.}~\bibnamefont
  {Rey}}, \bibinfo {author} {\bibfnamefont {L.}~\bibnamefont {Jiang}}, \bibinfo
  {author} {\bibfnamefont {M.}~\bibnamefont {Fleischhauer}}, \bibinfo {author}
  {\bibfnamefont {E.}~\bibnamefont {Demler}}, \ and\ \bibinfo {author}
  {\bibfnamefont {M.}~\bibnamefont {Lukin}},\ }\href@noop {} {\bibfield
  {journal} {\bibinfo  {journal} {Physical Review A}\ }\textbf {\bibinfo
  {volume} {77}},\ \bibinfo {pages} {052305} (\bibinfo {year}
  {2008})}\BibitemShut {NoStop}%
\bibitem [{\citenamefont {Cappellaro}\ and\ \citenamefont
  {Lukin}(2009)}]{cappellaro2009quantum}%
  \BibitemOpen
  \bibfield  {author} {\bibinfo {author} {\bibfnamefont {P.}~\bibnamefont
  {Cappellaro}}\ and\ \bibinfo {author} {\bibfnamefont {M.~D.}\ \bibnamefont
  {Lukin}},\ }\href {\doibase 10.1103/PhysRevA.80.032311} {\bibfield  {journal}
  {\bibinfo  {journal} {Phys. Rev. A}\ }\textbf {\bibinfo {volume} {80}},\
  \bibinfo {pages} {032311} (\bibinfo {year} {2009})}\BibitemShut {NoStop}%
\bibitem [{\citenamefont {Rudner}\ \emph {et~al.}(2011)\citenamefont {Rudner},
  \citenamefont {Vandersypen}, \citenamefont {Vuleti{\'c}},\ and\ \citenamefont
  {Levitov}}]{rudner2011generating}%
  \BibitemOpen
  \bibfield  {author} {\bibinfo {author} {\bibfnamefont {M.}~\bibnamefont
  {Rudner}}, \bibinfo {author} {\bibfnamefont {L.}~\bibnamefont {Vandersypen}},
  \bibinfo {author} {\bibfnamefont {V.}~\bibnamefont {Vuleti{\'c}}}, \ and\
  \bibinfo {author} {\bibfnamefont {L.}~\bibnamefont {Levitov}},\ }\href@noop
  {} {\bibfield  {journal} {\bibinfo  {journal} {Physical Review Letters}\
  }\textbf {\bibinfo {volume} {107}},\ \bibinfo {pages} {206806} (\bibinfo
  {year} {2011})}\BibitemShut {NoStop}%
\bibitem [{\citenamefont {Bouchoule}\ and\ \citenamefont
  {M\o{}lmer}(2002)}]{bouchoule2002spin}%
  \BibitemOpen
  \bibfield  {author} {\bibinfo {author} {\bibfnamefont {I.}~\bibnamefont
  {Bouchoule}}\ and\ \bibinfo {author} {\bibfnamefont {K.}~\bibnamefont
  {M\o{}lmer}},\ }\href@noop {} {\bibfield  {journal} {\bibinfo  {journal}
  {Phys. Rev. A}\ }\textbf {\bibinfo {volume} {65}},\ \bibinfo {pages} {041803}
  (\bibinfo {year} {2002})}\BibitemShut {NoStop}%
\bibitem [{\citenamefont {Gil}\ \emph {et~al.}(2014)\citenamefont {Gil},
  \citenamefont {Mukherjee}, \citenamefont {Bridge}, \citenamefont {Jones},\
  and\ \citenamefont {Pohl}}]{gil2014spin}%
  \BibitemOpen
  \bibfield  {author} {\bibinfo {author} {\bibfnamefont {L.}~\bibnamefont
  {Gil}}, \bibinfo {author} {\bibfnamefont {R.}~\bibnamefont {Mukherjee}},
  \bibinfo {author} {\bibfnamefont {E.}~\bibnamefont {Bridge}}, \bibinfo
  {author} {\bibfnamefont {M.}~\bibnamefont {Jones}}, \ and\ \bibinfo {author}
  {\bibfnamefont {T.}~\bibnamefont {Pohl}},\ }\href@noop {} {\bibfield
  {journal} {\bibinfo  {journal} {Physical Review Letters}\ }\textbf {\bibinfo
  {volume} {112}},\ \bibinfo {pages} {103601} (\bibinfo {year}
  {2014})}\BibitemShut {NoStop}%
\bibitem [{\citenamefont {Kaubruegger}\ \emph {et~al.}(2019)\citenamefont
  {Kaubruegger}, \citenamefont {Silvi}, \citenamefont {Kokail}, \citenamefont
  {van Bijnen}, \citenamefont {Rey}, \citenamefont {Ye}, \citenamefont
  {Kaufman},\ and\ \citenamefont {Zoller}}]{kaubruegger2019variational}%
  \BibitemOpen
  \bibfield  {author} {\bibinfo {author} {\bibfnamefont {R.}~\bibnamefont
  {Kaubruegger}}, \bibinfo {author} {\bibfnamefont {P.}~\bibnamefont {Silvi}},
  \bibinfo {author} {\bibfnamefont {C.}~\bibnamefont {Kokail}}, \bibinfo
  {author} {\bibfnamefont {R.}~\bibnamefont {van Bijnen}}, \bibinfo {author}
  {\bibfnamefont {A.~M.}\ \bibnamefont {Rey}}, \bibinfo {author} {\bibfnamefont
  {J.}~\bibnamefont {Ye}}, \bibinfo {author} {\bibfnamefont {A.~M.}\
  \bibnamefont {Kaufman}}, \ and\ \bibinfo {author} {\bibfnamefont
  {P.}~\bibnamefont {Zoller}},\ }\href@noop {} {\bibfield  {journal} {\bibinfo
  {journal} {Physical Review Letters}\ }\textbf {\bibinfo {volume} {123}},\
  \bibinfo {pages} {260505} (\bibinfo {year} {2019})}\BibitemShut {NoStop}%
\bibitem [{\citenamefont {Borish}\ \emph {et~al.}(2020)\citenamefont {Borish},
  \citenamefont {Markovi{\'c}}, \citenamefont {Hines}, \citenamefont
  {Rajagopal},\ and\ \citenamefont {Schleier-Smith}}]{borish2020transverse}%
  \BibitemOpen
  \bibfield  {author} {\bibinfo {author} {\bibfnamefont {V.}~\bibnamefont
  {Borish}}, \bibinfo {author} {\bibfnamefont {O.}~\bibnamefont
  {Markovi{\'c}}}, \bibinfo {author} {\bibfnamefont {J.~A.}\ \bibnamefont
  {Hines}}, \bibinfo {author} {\bibfnamefont {S.~V.}\ \bibnamefont
  {Rajagopal}}, \ and\ \bibinfo {author} {\bibfnamefont {M.}~\bibnamefont
  {Schleier-Smith}},\ }\href@noop {} {\bibfield  {journal} {\bibinfo  {journal}
  {Physical Review Letters}\ }\textbf {\bibinfo {volume} {124}},\ \bibinfo
  {pages} {063601} (\bibinfo {year} {2020})}\BibitemShut {NoStop}%
\bibitem [{\citenamefont {Mudry}\ and\ \citenamefont
  {Fradkin}(1989)}]{mudry1989ground}%
  \BibitemOpen
  \bibfield  {author} {\bibinfo {author} {\bibfnamefont {C.}~\bibnamefont
  {Mudry}}\ and\ \bibinfo {author} {\bibfnamefont {E.}~\bibnamefont
  {Fradkin}},\ }\href@noop {} {\bibfield  {journal} {\bibinfo  {journal}
  {Physical Review B}\ }\textbf {\bibinfo {volume} {40}},\ \bibinfo {pages}
  {11177} (\bibinfo {year} {1989})}\BibitemShut {NoStop}%
\bibitem [{\citenamefont {Luo}\ \emph {et~al.}(2017)\citenamefont {Luo},
  \citenamefont {Zou}, \citenamefont {Wu}, \citenamefont {Liu}, \citenamefont
  {Han}, \citenamefont {Tey},\ and\ \citenamefont
  {You}}]{luo2017deterministic}%
  \BibitemOpen
  \bibfield  {author} {\bibinfo {author} {\bibfnamefont {X.-Y.}\ \bibnamefont
  {Luo}}, \bibinfo {author} {\bibfnamefont {Y.-Q.}\ \bibnamefont {Zou}},
  \bibinfo {author} {\bibfnamefont {L.-N.}\ \bibnamefont {Wu}}, \bibinfo
  {author} {\bibfnamefont {Q.}~\bibnamefont {Liu}}, \bibinfo {author}
  {\bibfnamefont {M.-F.}\ \bibnamefont {Han}}, \bibinfo {author} {\bibfnamefont
  {M.~K.}\ \bibnamefont {Tey}}, \ and\ \bibinfo {author} {\bibfnamefont
  {L.}~\bibnamefont {You}},\ }\href@noop {} {\bibfield  {journal} {\bibinfo
  {journal} {Science}\ }\textbf {\bibinfo {volume} {355}},\ \bibinfo {pages}
  {620} (\bibinfo {year} {2017})}\BibitemShut {NoStop}%
\bibitem [{\citenamefont {Zhiqiang}\ \emph {et~al.}(2017)\citenamefont
  {Zhiqiang}, \citenamefont {Lee}, \citenamefont {Kumar}, \citenamefont
  {Arnold}, \citenamefont {Masson}, \citenamefont {Parkins},\ and\
  \citenamefont {Barrett}}]{zhiqiang2017nonequilibrium}%
  \BibitemOpen
  \bibfield  {author} {\bibinfo {author} {\bibfnamefont {Z.}~\bibnamefont
  {Zhiqiang}}, \bibinfo {author} {\bibfnamefont {C.~H.}\ \bibnamefont {Lee}},
  \bibinfo {author} {\bibfnamefont {R.}~\bibnamefont {Kumar}}, \bibinfo
  {author} {\bibfnamefont {K.}~\bibnamefont {Arnold}}, \bibinfo {author}
  {\bibfnamefont {S.~J.}\ \bibnamefont {Masson}}, \bibinfo {author}
  {\bibfnamefont {A.}~\bibnamefont {Parkins}}, \ and\ \bibinfo {author}
  {\bibfnamefont {M.}~\bibnamefont {Barrett}},\ }\href@noop {} {\bibfield
  {journal} {\bibinfo  {journal} {Optica}\ }\textbf {\bibinfo {volume} {4}},\
  \bibinfo {pages} {424} (\bibinfo {year} {2017})}\BibitemShut {NoStop}%
\end{thebibliography}%

\clearpage


\section*{Supplementary material}

# Protecting Spin Coherence in a Tunable Heisenberg Model: Supplemental Material

(Dated: August 30, 2020)

This supplement provides supporting derivations and details of the experimental methods. In Sec. I, we derive the implemented Hamiltonian and present a simple toy model for the protection of spin coherence. In Sec. II, we elaborate on the experimental sequences and calibrations of atom-cavity coupling and atom number. In Sec. III, we elaborate on the data analysis and modeling.

## I. THEORETICAL BACKGROUND

### A. Derivation of Tunable Heisenberg Hamiltonian

A theoretical derivation of the cavity-mediated Heisenberg Hamiltonian has been presented in Ref. [1], building on our implementation and analysis for the special case of the XY model in Ref. [2]. Here, we review and expand upon the derivation of the effective spin model

$$H_{\text{XXZ}} = J^{xy}(\Theta)\left(\mathcal{F}_x^2 + \mathcal{F}_y^2\right) + J^z(\Theta)\mathcal{F}_z^2, \tag{S1}$$

where $\mathcal{F}$ is the weighted collective spin defined in the main text. We focus on the relationship of the couplings $J^{xy,z}$ to experimental parameters.

The spin-spin interactions arise from a Faraday interaction between the atoms and the cavity field

$$H_I = \Omega\left(a_+^\dagger a_+ - a_-^\dagger a_-\right)\mathcal{F} \cdot \hat{\mathbf{z}}_c, \tag{S2}$$

where $a_\pm$ represent a pair of circularly polarized cavity modes with respect to the cavity axis $\hat{\mathbf{z}}_c$, and $\Omega$ is the average vector ac Stark shift per intracavity photon. Only the spin projection along the cavity axis $\hat{\mathbf{z}}_c$ interacts with the light field. For a cavity driven with linearly polarized light at frequency $\omega_d$, detuned by $\delta = \omega_d - \omega_0$ from cavity resonance, we can expand the atom-cavity interaction Hamiltonian to first order in fluctuations of the vertically polarized cavity mode $v$ to obtain:

$$H_I \approx \frac{\Omega}{\sqrt{(\kappa/2)^2 + \delta^2}}\left(\varepsilon e^{-i\delta t}v^\dagger + \text{h.c.}\right)\left[\mathcal{F}_z\cos\Theta + \frac{1}{2}\left(\mathcal{F}_+ e^{i\omega_z t} + \mathcal{F}_- e^{-i\omega_z t}\right)\sin\Theta\right], \tag{S3}$$

where $\varepsilon$ parameterizes the drive strength. Provided the occupation of the $v$ mode remains small, we can adiabatically eliminate it from the dynamics to arrive at an effective Hamiltonian for the spins:

$$H_{\text{eff}} = \Omega^2 n\left[\mathcal{F}_z\mathcal{F}_z\ \xi(\delta)\cos^2\Theta + \frac{1}{4}\mathcal{F}_+\mathcal{F}_-\ \xi(\delta_-)\sin^2\Theta + \frac{1}{4}\mathcal{F}_-\mathcal{F}_+\ \xi(\delta_+)\sin^2\Theta\right] \tag{S4}$$

where $n = |\varepsilon|^2/[\delta^2 + (\kappa/2)^2]$ is the average intracavity photon number, $\delta_\pm = \delta \mp \omega_Z$ are the detunings from Raman resonances at $\delta = \pm\omega_Z$, and $\xi(\delta) = \delta/[\delta^2 + (\kappa/2)^2]$.

The effective Hamiltonian $H_{\text{eff}}$ in Eq. S4 is related to $H_{\text{XXZ}}$ by

$$H_{\text{eff}} = H_{\text{XXZ}} + \frac{1}{4}\Omega^2 n\sum_i c_i^2 f_i^z\left[\xi(\delta_-)) - \xi(\delta_+)\right] \tag{S5}$$

where the second term is negligible for large $\mathcal{F}$ and vanishes in the limit of large detuning $\delta_\pm \gg \omega_Z$. For arbitrary detuning, Eq. S4 shows that the couplings $J^{xy,z}$ take the form

$$J^{xy} = n\Omega^2 \mathcal{A}^{x,y}(\Theta, \delta) \tag{S6a}$$

$$J^z = n\Omega^2 \mathcal{A}^z(\Theta, \delta). \tag{S6b}$$

Here, the dependence on tuning angle $\Theta$ and detuning $\delta$ is captured by the functions

$$\mathcal{A}^z(\Theta, \delta) = \xi(\delta)\cos^2\Theta, \tag{S7a}$$

$$\mathcal{A}^{xy}(\Theta, \delta) = \frac{\sin^2\Theta}{4}\left[\xi(\delta_-) + \xi(\delta_+)\right]. \tag{S7b}$$

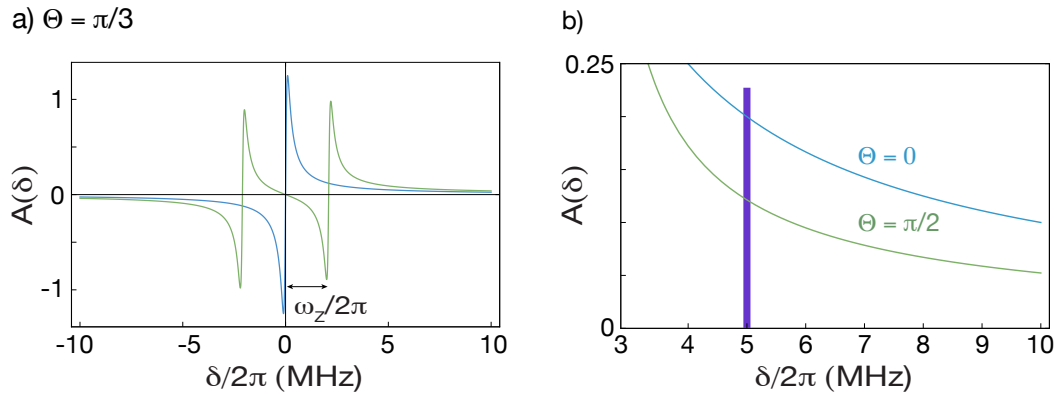

FIG. S1: Dependence of interactions on detuning for $(\kappa, \omega_Z) = 2\pi \times (0.2, 2.1)$ MHz. (a) The functions $\mathcal{A}^{xy}(\Theta = \pi/3, \delta)$ (green) and $\mathcal{A}^z(\Theta = \pi/3, \delta)$ (blue) are plotted here at fixed $\Theta$ to show that all four relative signs of XY and Ising couplings are accessible. (b) We zoom in on the functions $\mathcal{A}^{xy}(\Theta = \pi/2, \delta)$ (green) and $\mathcal{A}^z(\Theta = 0, \delta)$ (blue), which determine the strengths of XY and Ising couplings, at larger detuning. In the large detuning limit, $\mathcal{A}^z(0, \delta)/\mathcal{A}^{xy}(\pi/2, \delta) = 2$. Our drive field (purple) is typically placed around $\delta/2\pi \sim 5$ MHz, where this ratio is 1.6.

They are plotted for $\Theta = \pi/3$ in Fig. S1a, illustrating that all four relative signs $(++), (--), (+-), (-+)$ of the Ising and XY couplings are achievable by adjusting the drive detuning. The couplings at large detuning $\delta \gg \kappa, \omega_Z$ compared to the cavity linewidth $\kappa$ and Zeeman splitting $\omega_Z$ are plotted in Fig. S1b. In this regime, Eq. S5 simplifies to

$$H_{\text{eff}} = \frac{n\Omega^2}{\delta}\left[\cos^2\Theta \mathcal{F}_z^2 + \frac{1}{2}\sin^2\Theta\left(\mathcal{F}_x^2 + \mathcal{F}_y^2\right)\right], \tag{S8}$$

and $J_z(0)/J^{xy}(\pi/2) = 2$. For the drive detunings $\delta \approx 2\pi \times 5$ MHz used for the measurements in Fig. 2c of the main text, we expect a ratio of 1.6, consistent with the displayed fits.

### B. Model for Dephasing

Here we present an estimate of the critical interaction strength for protecting spin coherence in a simplified system. We consider a collection of atoms in which the only source of inhomogeneity is a field along $\hat{z}$ that takes on values $+h_z\hat{z}$ and $-h_z\hat{z}$ in two different regions of the cloud. We assume that the collective spin vectors $\mathbf{S}_1$ and $\mathbf{S}_2$ in these two regions are both initially polarized along $\hat{x}$ and analyze their response to the inhomogeneous field. The total Hamiltonian is

$$H_{\text{XXZ}}/\hbar + H_{\text{inh}}/\hbar = J^{xy}(\Theta)\left[F_x^2 + F_y^2\right] + J^z(\Theta)F_z^2 + \sum_i h_{i,z} S_{i,z}, \tag{S9}$$

where $\mathbf{F} = \sum_i \mathbf{S}_i$ is the equally weighted sum of the two collective spin vectors. $F_z$ commutes with the Hamiltonian and is thus conserved. The system is also invariant under exchange of the two spins followed by a 180 degree rotation about $\hat{x}$. Due to this symmetry, we can describe the whole system with the three components $(S_x, S_y, S_z)$ of a single collective spin. Ignoring quantum fluctuations, the other spin has components $(S_x, -S_y, -S_z)$ and the total spin is $\mathbf{F} = (2S_x, 0, 0)$.

The dynamics of the spin $\mathbf{S}$ are constrained by two more conserved quantities: the length of each collective spin and the total energy of the system. The equation for each collective spin, $S^2 = S_x^2 + S_y^2 + S_z^2 = \text{const}$, maps out the surface of a Bloch sphere. The total energy is given by the Hamilitonian, which using the classical expression for the total spin reduces to $4J^{xy}(\Theta)S_x^2 + 2h_z S_z = E$.

We plot projections of constant energy contours on the surface of the Bloch sphere in the equatorial plane in Fig. S2. These contours are qualitatively different for values of $J^{xy}$ above and below the critical point in this model, $J_c^{xy} = 2h_z/S$. When $SJ^{xy} < 2h_z$, the contours encircle the north pole of the Bloch sphere, permitting dephasing. For larger values of $J^{xy}(\Theta)$, the contours break into two disjoint cycles and the two collective spins remain localized,

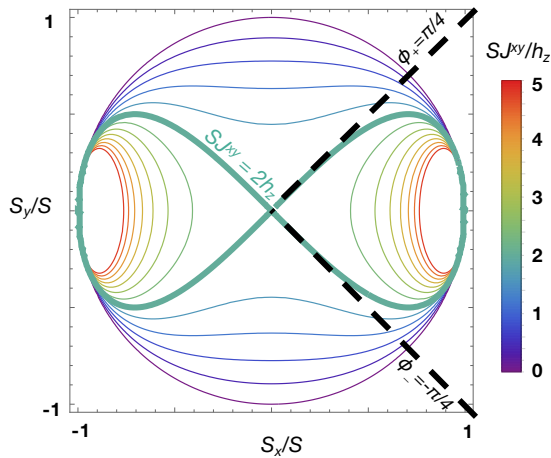

FIG. S2: **Toy model for critical interaction strength.** Contours of constant energy are sketched on the Bloch sphere as viewed from the north pole, for spins initially aligned with $\hat{\mathbf{x}}$. The critical interaction strength (thick green curve) is labeled. Above this critical strength, the phase evolution of a single spin ($\mathbf{S}_1$ or $\mathbf{S}_2$) along a constant-energy contour remains between $\phi_\pm = \pm\pi/4$, bounding the angle between spins $\mathbf{S}_1$ and $\mathbf{S}_2$ to be at most $\phi_+ - \phi_- = \pi/2$.

.

bounding the phase accumulation between them. This approach agrees with the scaling of the estimate based on the energy gap.

## II. EXPERIMENTAL METHODS

### A. Preparation and Detection

The atoms are cooled and trapped in a magneto-optical trap in the center of the cavity, and then loaded into the trapping lattice after an optical molasses cooling stage. In lieu of optical pumping, atoms not in $|f=1, m_f=-1\rangle$ are removed by adiabatically sweeping $m_f = 0, 1$ atoms into the $f = 2$ manifold, then pushing them out of the trap with resonant light on the $\left|5S_{1/2}, f = 2\right\rangle \to \left|5P_{3/2}, f' = 3\right\rangle$ transition. We thus obtain a spin-polarized state in $|f=1, m_f=-1\rangle$, which forms the starting point for preparing any desired spin texture using the focused scanning Raman beam (Fig. S3). After state preparation, the magnetic field is tuned to the angle $\Theta$, and the drive laser is then switched on to generate interactions. Throughout the interaction time, the trap depth is $h \times 5$ MHz corresponding to a transverse trap frequency of 1.5 kHz.

We perform state detection by measuring the three spin components ($\langle f_x \rangle, \langle f_y \rangle, \langle f_z \rangle$) in three separate runs of the experiment. In order to measure the magnetization $\langle f_z \rangle \equiv \rho_1 - \rho_{-1}$, where $\rho_k$ is the local atomic density in the $m_f = k$ Zeeman level, we perform state-sensitive imaging in each of the three Zeeman levels $m_f = \pm 1, 0$. Sequentially, we transfer each $m_f$ population into $f = 2$ with an adiabatic microwave sweep and perform fluorescence imaging on the $\left|5S_{1/2}, f = 2\right\rangle \to \left|5P_{3/2}, f' = 3\right\rangle$ transition, simultaneously lowering the trap so the imaged atoms are kicked out of the trap by the imaging light. In order to measure the transverse spin components $\langle f_x \rangle$ or $\langle f_y \rangle$, we image after an additional global Raman $\pi/2$ pulse with phase 0 or $\pi/2$, respectively, to rotate the transverse spin component into the $z$-basis.

We calibrate our imaging pulses by measuring Rabi oscillations induced by a global Raman beam. The distribution of atoms in the three $m_f$ levels during these oscillations reveals systematic errors including atom loss on each imaging pulse and imperfect removal of previously imaged states, as well as residual atoms initially in $m_f \neq -1$. To characterize these errors, we optimize a $3 \times 3$ matrix $M$ to minimize $\sum_t (M\boldsymbol{\rho}^{\text{meas}}(t) - \boldsymbol{\rho}^{\text{theory}}(t))^2$. Here, $\boldsymbol{\rho}$ is a vector specifying the populations in the three Zeeman states and $\boldsymbol{\rho}^{\text{theory}}$ describes Rabi oscillations of full contrast, as justified by the measured Rabi coherence time. Assuming that the imperfections in imaging and state preparation are constant for a given data run, we then apply $M$ on all measured data to accurately determine the magnetization.

## B. Atom-Cavity Coupling

Here we derive the dimensionless weights $c(\zeta)$ parameterizing the spatial dependence of the Faraday coupling of atoms to the cavity mode. These weights depend both on the position $\zeta = z_c/z_R$ at which the atoms are pinned by the 1560 lattice and on the temperature, which determines the radial and axial distributions of atoms within each lattice site.

As a function of distance from cavity center $z_c$ and radial distance $r$ from the optical axis, the intensity profiles (neglecting wavefront curvature) for the two modes are given by

$$I/I_{\max} = (w_0/w)^2 e^{-2r^2/w^2} \cos^2(kz_c - \psi), \tag{S10}$$

where $\psi(\zeta) = \tan^{-1}(\zeta)$ is the Gouy phase shift and $w(\zeta) = w_0\sqrt{1+\zeta^2}$ is the waist. The Rayleigh range $z_R = 1.4$ mm is determined by the cavity geometry and is the same for both lattices. The wavenumbers and waists for the two lattices are related by $k_{780} = 2k_{1560}$ and $w_{780}(\zeta) = w_{1560}(\zeta)/\sqrt{2}$.

Atoms imaged at position $\zeta$ are trapped around the nearest antinode of the 1560 nm lattice, $z_{peak} = (n\pi + \psi(\zeta))/k_{1560}$. Approximating the trap potential as harmonic near the antinode, the thermal distribution of atoms is

$$P(\Delta z, r) \propto e^{-(k_{1560}\Delta z)^2/\tau} e^{-2(r/w_{1560})^2/\tau}, \tag{S11}$$

where $\Delta z \equiv z_c - z_{peak}$ and $\tau$ is the temperature normalized to trap depth at a given position $\zeta$. For the same trapping site at position $\zeta$, the intensity of the 780 nm drive mode is

$$I_{780}(\Delta z, r) = \frac{I_{\max}}{1+\zeta^2} e^{-2r^2/w_{780}^2} \cos^2(k_{780}\Delta z + \psi(\zeta)). \tag{S12}$$

We compute the couplings $c(\zeta)$, normalized to the coupling $c_{\max}$ of an atom at cavity center ($z_c = r = 0$), by averaging the intensity of the light field over the distribution of atoms,

$$\frac{c(\zeta)}{c_{\max}} = \frac{\int I_{780}(\Delta z, r) P(\Delta z, r) d^3r}{\int I_{\max} P(\Delta z, r) d^3r} = \frac{1}{1+2\tau}\left[\frac{e^{-4\tau}}{(1+\zeta^2)^2} + \frac{1}{2}\frac{1-e^{-4\tau}}{1+\zeta^2}\right]. \tag{S13}$$

The factor $1/(1+2\tau)$ comes from comparing the radial distribution to the waist of the drive mode while $e^{-4\tau}$ compares the width of the axial distribution to the lattice spacing. When atoms are warm, they experience an averaged probe coupling that decays as $w^{-2}$. When the atoms are cold, coupling to the drive mode at cavity center is higher, but the offset between the maxima of the trapping lattice and the maxima of the drive mode (due to the Gouy phase shift) causes coupling to decay more quickly as $w^{-4}$.

The functional form of the couplings in Eq. S13 is corroborated by spectroscopic measurements of the vector light shift, similar to the measurement described in Ref. [2]. Fitting the spatial dependence of the vector light shift yields a temperature parameter $\tau = 0.3(1)$. When the couplings are normalized such that their average value is 1, the maximum coupling becomes $c_{\max} = \Omega_0/\Omega$. We compute $\Omega_0$ from $g$ as in the main text. For a maximally coupled atom at cavity center, $g^2 = 6\Gamma\omega_{\text{FSR}}/(\pi k_{780}^2 w_0^2)$ where $\Gamma = 2\pi \times 6.07$ MHz is the linewidth of the transition and $\omega_{\text{FSR}} = 2\pi \times 3.0$ GHz is the free spectral range of the cavity [3]. The waist of the 780 nm mode is $w_0 = 18.6$ $\mu$m, calculated from the measured transverse mode spacing. These values yield $2g = 2\pi \times 2.5$ MHz corresponding to $\Omega_0 = 2\pi \times 23$ Hz. We then determine $\Omega$ using Eq. S13 with our measured value for $\tau$.

## C. Cavity Shift and Atom Number

We calibrate the atom number by measuring the dispersive shift of the cavity resonance frequency, which is proportional to the total weighted atom number $N$. In particular, in addition to the Faraday interaction that generates birefringence proportional to $\mathcal{F}_z$, the Hamiltonian includes a term

$$H_s = 4\Omega N \left(a_+^\dagger a_+ + a_-^\dagger a_-\right). \tag{S14}$$

This term describes a common-mode shift of the $\sigma_+$ and $\sigma_-$ cavity resonances that depends on the number of coupled atoms

$$\Delta\omega \equiv \omega_0(N) - \omega_0(0) = -4\Omega N. \tag{S15}$$

In the main text, we let $\omega_0 \equiv \omega_0(N)$ represent the atom-shifted cavity resonance frequency for total weighted atom number $N$. For the measurements in the main text, we typically measure $\Delta\omega \approx -2\pi \times 2.5$ MHz. In combination with our determination of the average single-photon vector light shift $\Omega$ (Sec. II B), we use $\Delta\omega$ to calibrate the total atom number, and to calibrate the number of counts per atom obtained in images of the magnetization.

### D. Experimental Sequences

In this section, we give further details describing the experimental sequences used throughout the main text. In particular, we give numerical values for relevant experimental parameters and provide qualitative illustrations of the pulse sequences and time-changing parameters for conceptual clarity.

**Hamiltonian tomography.** The sequences used for Hamiltonian tomography as shown in Fig. 2 of the main text are illustrated in Fig. S3. These measurements were performed in a 3 G field. To measure the Ising coupling or XY couplings, the state $|\psi_z\rangle$ or $|\psi_y\rangle$ is prepared via both local and global Raman rotations. The weighted probe spins in regions $B$ and $C$ point in opposite directions and are equal in length to within 5% of the total weighted spin vector. After preparation of the target initial state, the drive laser is then switched on for $t_{\text{int}}$ to generate interactions. A final global $\pi/2$ pulse with phase $\varphi = 0$ ($\pi/2$) then rotates the desired quadrature $x$ ($y$) into the population basis, whence it is measured with the detection sequence described in Sec. II A. Alternatively, $\langle f_z \rangle$ can be measured directly without the extra $\pi/2$ rotation.

For measurements of the Ising coupling, an optional spin echo may be performed by adding a global $\pi$ pulse halfway through the interaction time (Fig. S3a). The spin echo ensures that the measured spin precession is solely due to interactions, and would cancel the effect of any circular polarization of the intracavity light that is independent of the atomic magnetization. In the main text, all measurements of $J^{xy,z}(\Theta)$ in Fig. 2(c) were performed with spin echo, and are consistent with the measurement in Fig. 2(a) that is performed without spin echo.

The location of the atom cloud shifted between measurements of the XY and Ising couplings, resulting in slightly different optimal choices for the regions $A, B, C$.

**Magnetic susceptibility.** The sequence for measuring susceptibility as in Fig. 3 of the main text is illustrated in Fig. S4a. We prepare the ground state of an effective field $H_0 = h_x F_x + h_z F_z$. The interactions are then ramped on, thus preparing a low-energy state of $H_{\text{tot}} = H_{\text{XXZ}} + H_0$ (but not the quantum ground state). Finally, the effective field and interactions are simultaneously quenched off and $\langle f_z \rangle$ is measured immediately. These measurements were performed in magnetic fields ranging between 1 and 3 G.

**Dephasing and gap protection.** The sequence for measuring dephasing as in Fig. 4 of the main text is illustrated in Fig. S4b. We prepare the ground state of $H_0 = h_x F_x + \sum_i h_{i,z} f_{i,z}$, where $h_{i,z}$ varies linearly across the cloud as $h_z(\zeta) = \mu\zeta$. The interactions are then ramped on, and the spins evolve under $H_{\text{XXZ}} + \sum_i h_{i,z} f_{i,z}$ for a time $0 \leq t_{\text{wait}} \leq 1$ ms before the measurement of $\langle f_x \rangle, \langle f_y \rangle, \langle f_z \rangle$. These measurements were performed in a 1 G field.

**Experimental timescales.** In each of the measurement sequences described above, the choice of overall timescale for the experiment is governed by technical considerations. The timescale for the experiments in Fig. 2 is lower-bounded by our timing resolution and the cavity lifetime, i.e. we cannot resolve time steps smaller than a few microseconds. The experiment length is upper-bounded by several factors, including inhomogeneous broadening due to a residual external magnetic field gradient and transverse atomic motion. We chose the timescale for the data in Fig. 2 to lie between these two limits. In Figs. 3 and 4, the timescale is somewhat longer and lower-bounded by the Rabi frequency of the Raman coupling and the magnetic field gradient.

### III. ANALYSIS AND MODELING

#### A. Mean-Field Simulations of Dynamics under $H_{\text{XXZ}}$

To verify our understanding of the dynamics used for Hamiltonian tomography, we simulate the evolution under $H_{\text{XXZ}}$ in a mean-field approximation. The mean-field dynamics are obtained by computing the Heisenberg equations of motion $dO/dt = i[H, O]$ and substituting $O \to \langle O \rangle$, thus ignoring all higher-order correlations. Concretely, we numerically simulate a system with $\beta$ spatial modes along the cavity axis by solving $3\beta$ coupled differential equations

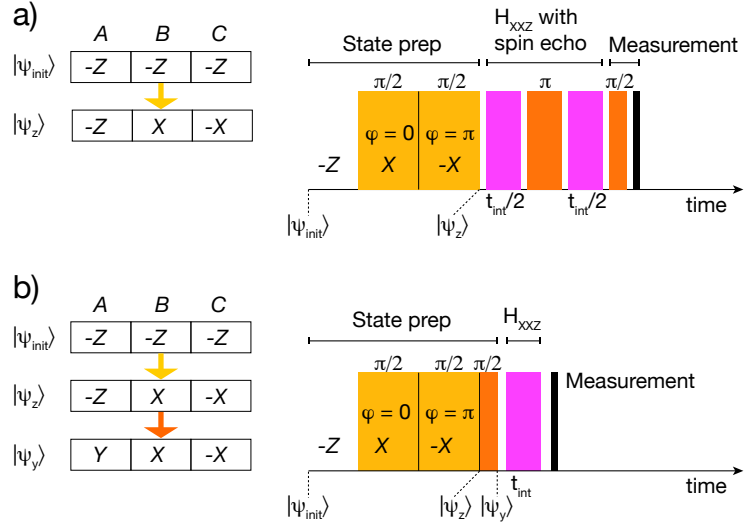

FIG. S3: Sequences for measuring Ising and XY Couplings. (a) To measure the Ising coupling, $|\psi_z\rangle$ is prepared by scanning a focused Raman beam of variable amplitude and phase (yellow) across the cloud. Locally, it performs a $\pi/2$ rotation with the desired phase such that the spins point along $\hat{\mathbf{x}}$ ($\varphi = 0$) or $-\hat{\mathbf{x}}$ ($\varphi = \pi$) in regions $B$ and $C$. To leave the atoms in region $A$ pointing down along $-\hat{\mathbf{z}}$, the Raman amplitude is locally set to zero. The drive laser is then switched on for a time $t_{\text{int}}$ (pink) to generate interactions. The optional spin echo consists of a $\pi$ pulse applied with a global Raman beam (orange) in the middle of $t_{\text{int}}$. A final global $\pi/2$ pulse (orange) then rotates the desired quadrature $x$ ($y$) into the population basis and the $m_f = \pm 1, 0$ populations are measured (black). (b) To measure the XY coupling, the state $|\psi_y\rangle$ is prepared by making $|\psi_z\rangle$ as in (a). The $-\hat{\mathbf{z}}$ spins are then rotated by a global $\pi/2$ pulse about $-\hat{\mathbf{x}}$ (orange) to point along $-\hat{\mathbf{y}}$. The drive laser is switched on for time $t_{\text{int}}$, after which the populations in $m_f = \pm 1, 0$ are measured (black).

for $(\langle f_x\rangle, \langle f_y\rangle, \langle f_z\rangle)$:

$$\frac{d\langle f_x(t,\zeta)\rangle}{dt} = -2c(\zeta)\left[J^z \mathcal{F}_z(t)\langle f_y(t,\zeta)\rangle - J^{xy}\mathcal{F}_y(t)\langle f_z(t,\zeta)\rangle\right] \quad \text{(S16a)}$$

$$\frac{d\langle f_y(t,\zeta)\rangle}{dt} = 2c(\zeta)\left[J^z \mathcal{F}_z(t)\langle f_x(t,\zeta)\rangle - J^{xy}\mathcal{F}_x(t)\langle f_z(t,\zeta)\rangle\right] \quad \text{(S16b)}$$

$$\frac{d\langle f_z(t,\zeta)\rangle}{dt} = -2J^{xy}c(\zeta)\left[\mathcal{F}_y(t)\langle f_x(t,\zeta)\rangle - \mathcal{F}_x(t)\langle f_y(t,\zeta)\rangle\right] \quad \text{(S16c)}$$

After plugging in measured values for the relevant quantities $\Omega, c(\zeta), \Delta\omega, \delta, \Theta$, we obtain the dynamics plotted in Fig. S5. These simulated dynamics, with the intracavity photon number $n$ as the only free parameter, are in good agreement with the experimental data in Fig. 2 of the main text. Similar equations are used to generate the mean-field model curves in Fig. 4d of the main text.

## B. Model for Susceptibility

We compare the measured magnetic susceptibility in Fig. 3 of the main text with a mean-field model, in which we approximate the collective spin vectors $\mathcal{F}$ and $\mathbf{F}$ by their expectation values. Since the system is initialized in a spin-polarized state, the weighted and unweighted collective spins initially have the same orientation $\mathcal{F} \parallel \mathbf{F}$. Assuming that this condition remains true throughout the susceptibility measurement, the XXZ Hamiltonian with additional effective fields

$$H_{\text{tot}} = J^z_{\text{eff}} \mathcal{F}_z^2 + h_x F_x + h_z F_z \quad \text{(S17)}$$

can be expressed in terms of the angle $\alpha \equiv \arctan(F_z/F_x)$ of the Bloch vector from the $\hat{\mathbf{x}}$-axis as

$$H_{\text{tot}} = J^z_{\text{eff}} \mathcal{F}^2 \sin^2\alpha + h_x F \cos\alpha + h_z F \sin\alpha. \quad \text{(S18)}$$

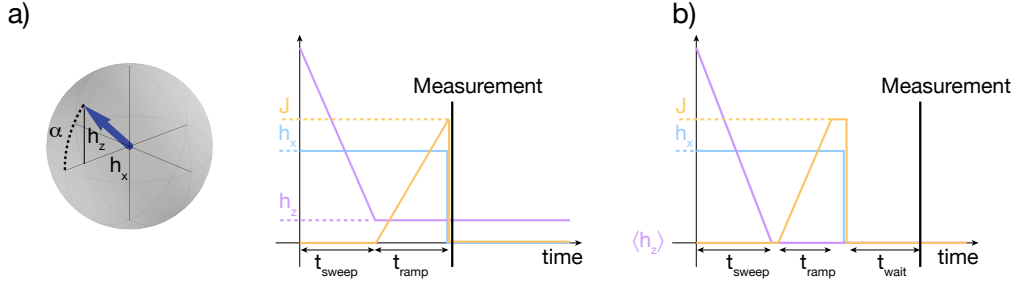

FIG. S4: Sequences for susceptibility and dephasing measurements. (a) To measure the magnetic susceptibility, we sweep the Raman detuning (purple) over a time $t_{\text{sweep}} = 5$ ms from far off-resonance to a final value $h_z$. At fixed Rabi frequency $h_x$ (blue), the angle $\alpha = \arctan(h_z/h_x)$ of the collective spin vector on the Bloch sphere, measured from the $\hat{\mathbf{x}}$-axis, is thus determined. The interactions (yellow) are then ramped on over a time $t_{\text{ramp}} = 5$ ms to a final value $J$, after which the effective field and interactions are simultaneously quenched off and the magnetization is measured immediately. (b) To measure the dephasing in a magnetic field gradient, the average Raman detuning (purple) is swept during $t_{\text{sweep}} = 5$ ms from far off-resonance to zero, such that the collective spin vector points on average along $\hat{\mathbf{x}}$. Interactions are then ramped on to a final value $J$ during $t_{\text{ramp}} = 5$ ms. The spins evolve under $H_{\text{XXZ}} + H_{\text{inh}}$ for a time $0 \leq t_{\text{wait}} \leq 1$ ms before the measurement is performed.

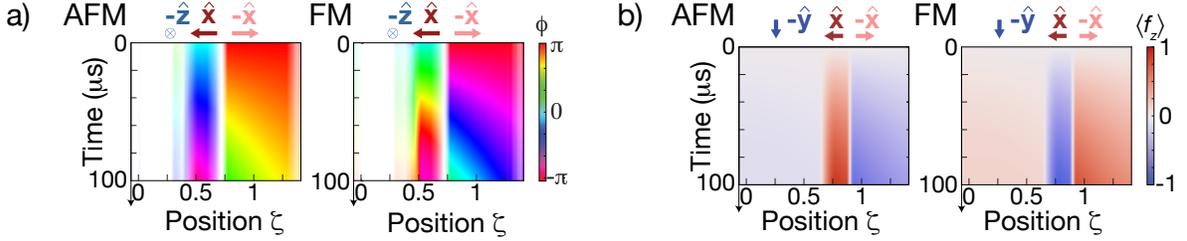

FIG. S5: Mean-field simulations corresponding to Fig. 2 of the main text for (a) Ising and (b) XY couplings. With $\Theta \approx 53°$, we plot (a) phase $\phi$, with opacity indicating transverse spin length, or (b) magnetization $\langle f_z \rangle$. We compare both signs of drive detuning $\delta$ to check the agreement for antiferromagnetic (AFM) [(a) $\delta = 2\pi \times 7.5$ MHz, (b) $2\pi \times 5.5$ MHz] and ferromagnetic (FM) [(a) $\delta = -2\pi \times 5.5$ MHz (b) $-2\pi \times 5.5$ MHz] couplings. The intracavity photon number varies from $n \approx 5000$ in (b) to $n \approx 15000$ in (a) AFM and 20000 in (a) FM, which agrees within uncertainty with the measured intracavity photon number for these data.

To find the minimum energy configuration for the angle $\alpha$ as a function of the parameters $h_x, h_z$, we solve

$$\frac{\partial H_{\text{tot}}}{\partial \alpha} = 2 J_{\text{eff}}^z \mathcal{F}^2 \sin\alpha \cos\alpha - h_x F \sin\alpha + h_z F \cos\alpha = 0. \tag{S19}$$

Taking the derivative of Eq. S19 with respect to $h_z$, we solve for the susceptibility, with $\alpha(h_z = 0) = 0$:

$$\chi \equiv \left.\frac{\partial \sin\alpha}{\partial (h_z/h_x)}\right|_{h_z=0} = \frac{\cos^2\alpha}{2 J_{\text{eff}}^z \mathcal{F}^2 (2\cos^2\alpha - 1)/h_x F + \cos\alpha} = \frac{1}{2 J_{\text{eff}}^z \mathcal{F}^2 / h_x F + 1} \tag{S20}$$

$$= \frac{1}{2\Lambda_{\text{eff}}^z/h_x + 1}, \tag{S21}$$

Note that in the main text we equivalently define the susceptibility in terms of the polar angle $\theta = \pi/2 - \alpha$.

Equation S21 shows that the susceptibility diverges at a critical value $\Lambda_{\text{eff}}^z = -h_x/2$ of the collective interaction parameter. The physical meaning of the diverging susceptibility at the critical point can be understood by examining the energy landscape (Eq. S19) above and below the critical interaction strength, as sketched in Fig. S6. For no interactions, the minimum energy shifts smoothly as a function of $h_z$, but at the critical point the parabolic landscape

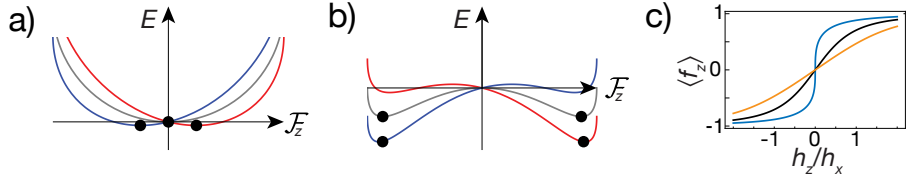

FIG. S6: (a-b) Energy landscape of transverse-field Ising model in the classical limit for three different values of $h_z$: $h_z < 0$ (red), $h_z > 0$ (blue), and $h_z = 0$ (gray). The energy minimum for the non-interacting model (a) varies smoothly with varying $h_z$. With ferromagnetic Ising interactions (b), the energy minimum instead jumps discontinuously between the two local minima of a tilted double-well potential. (c) Sample curves of the magnetic susceptibility as a function of $h_z/h_x$, plotted for $\Lambda^z_{\text{eff}}/h_x = -0.5$ (blue), $\Lambda^z_{\text{eff}}/h_x = 0$ (black), and $\Lambda^z_{\text{eff}}/h_x = 0.5$ (orange).

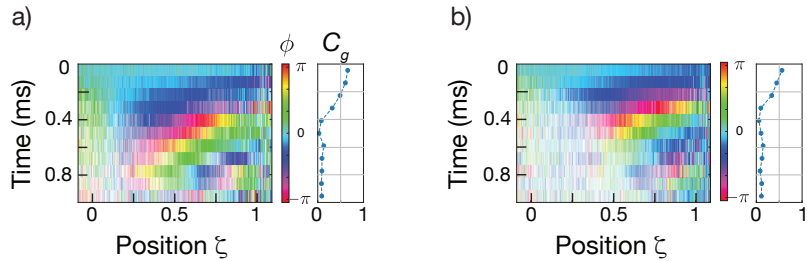

FIG. S7: (a-b) Phase $\phi$ and global contrast $C_g$ at $\Theta = 0$ for (a) no interactions and (b) Ising interactions $[J^z \mathcal{F}/\mu L = 0.52(5)]$, in magnetic field gradient $|\mu| = 2\pi \times 2$ kHz$/L$. Opacity indicates length of transverse spin component.

becomes a tilted double well potential, whose minimum jumps between two separate wells as $h_z$ is tuned through zero.

We use the model in Eq. S21 to fit the susceptibility measurements in Fig. 3(c) of the main text. There, the susceptibility $\chi$ is determined from measurements of the magnetization $\langle f_z \rangle$ as a function of $h_z/h_x$. Specifically, we fit the magnetization data using the the model in Eq. S19 for the angle $\alpha$ as a function of $J^z_{\text{eff}}, h_x, h_z$, which can in general be solved numerically to yield curves of the form shown in Fig. S6(c). We calibrate $h_x$ using a fit to the magnetization of the non-interacting system, where $\alpha = \arctan(h_z/h_x)$ [Fig. 3(b.i) of main text]. We then fit Eq. S19 numerically to the data with interactions, such as those shown in Fig. 3(b.ii-iii) of the main text, and determine the magnetic susceptibility $\chi$ from the slope of each fitted curve at $h_z = 0$. The results are plotted in Figs. 3c and 3d of the main text, showing $\chi$ as a function of $\Lambda^{xy,z}$ and of $\Theta$.

## C. Dephasing Analysis

In addition to the measurements at $\Theta = \pi/2$ shown in Fig. 4(a-b) of the main text, we also measure dephasing at $\Theta = 0$ with and without Ising interactions, as plotted in Fig. S7. We use the data with no interactions to calibrate the gradient across a region of length $L = 1$ mm by performing a linear fit to the phase $\phi_L(t)$. In Fig. 4c of the main text, we plot this fit line together with the phase $\phi_L(t)$ extracted from the data with Ising interactions.

[1] G. Bentsen, I.-D. Potirniche, V. B. Bulchandani, T. Scaffidi, X. Cao, X.-L. Qi, M. Schleier-Smith, and E. Altman, Phys. Rev. X **9**, 041011 (2019), URL https://link.aps.org/doi/10.1103/PhysRevX.9.041011.
[2] E. J. Davis, G. Bentsen, L. Homeier, T. Li, and M. H. Schleier-Smith, Phys. Rev. Lett. **122**, 010405 (2019), also see Supplemental Material., URL https://link.aps.org/doi/10.1103/PhysRevLett.122.010405.

[3] H. Tanji-Suzuki, I. D. Leroux, M. H. Schleier-Smith, M. Cetina, A. T. Grier, J. Simon, and V. Vuletić, in Advances in atomic, molecular, and optical physics (Elsevier, 2011), vol. 60, pp. 201–237.



\end{document}